\documentclass[11pt,a4paper]{article}
\usepackage[cp1251]{inputenc} %Windows

\usepackage{amsmath,amsthm,amssymb,epsfig,latexsym}
\usepackage{ulem}
\usepackage[english]{babel}
 \usepackage{color}

\def\rvec{|0\>}

\def\beq#1{\begin{equation}\label{#1}}
\def\ba#1{\begin{multline}\label{#1}}
\def\eeq{\end{equation}}
\def\ea{\end{multiline}}

\def\Izer{{\sf K}}

\def\<{\langle}
\def\>{\rangle}
\def\CC{{\mathbb C}}

\def\ot{\otimes}

\def\sk#1{\left(#1\right)}

\def\prodl{\mathop{\overleftarrow\prod}\limits}

\def\ct{\mathbb{T}}

\def\bbb{\mathbb{B}}

\def\ccR{\mathbb{R}}

\def\LL{{T}}

\def\la{u}
\def\mu{v}

\newcommand{\bla}{\bar u}
\newcommand{\bmu}{\bar v}
\def\as{\lambda}
\def\fr{{\mathsf r}}

\newcommand{\wb}[1]{\overline{#1}}
\newcommand{\nx}{n_x}

\newtheorem{prop}{Proposition}[section]
\newtheorem{lemma}{Lemma}[section]
\newtheorem{cor}{Corollary}[section]

\newcommand{\sth}{\scriptscriptstyle \rm I\hspace{-1pt}I\hspace{-1pt}I}

 \makeatletter
 \@addtoreset{equation}{section}
 \makeatother

\def\Yan{\cY(\mathfrak{gl}_3)}
\def\rvec{|0\>}

\def\Izer{{\sf K}}

\def\<{\langle}
\def\>{\rangle}
\def\CC{{\mathbb C}}

\def\ot{\otimes}

\def\sk#1{\left(#1\right)}

\def\prodl{\mathop{\overleftarrow\prod}\limits}

\def\ct{\mathbb{T}}

\def\bbb{\mathbb{B}}

\def\ccR{\mathbb{R}}

\def\LL{{T}}

\def\la{u}
\def\muu{v}

\def\as{\lambda}

\newcommand{\so}{\scriptscriptstyle \rm I}
\newcommand{\st}{\scriptscriptstyle \rm I\hspace{-1pt}I}
\newcommand{\stt}{\scriptscriptstyle \rm I\hspace{-1pt}I\hspace{-1pt}I}
\newcommand{\sttt}{\scriptscriptstyle \rm I\hspace{-1pt}V}

      \def\cU{{\cal U}}
\def\cV{{\cal V}}      
\def\cY{{\cal Y}}   

\newcommand{\mb}[1]{\quad\mbox{#1}\quad}
\newcommand{\ben}{\begin{eqnarray}}
\newcommand{\een}{\end{eqnarray}}
\newcommand{\nonu}{\nonumber\\}

\newcommand{\vph}{\varphi}

\begin{document}

\renewcommand*{\thefootnote}{\fnsymbol{footnote}}
\begin{flushright}
LAPTH-044/12
\end{flushright}

\vspace{20pt}

\begin{center}
\begin{LARGE}
{\bf Bethe vectors of $GL(3)$-invariant integrable models}
\end{LARGE}

\vspace{40pt}

\begin{large}
{S.~Belliard${}^a$, S.~Pakuliak${}^b$, E.~Ragoucy${}^c$, N.~A.~Slavnov${}^d$\footnote{samuel.belliard@univ-montp2.fr, pakuliak@theor.jinr.ru, eric.ragoucy@lapp.in2p3.fr, nslavnov@mi.ras.ru}}
\end{large}

 \vspace{12mm}

${}^a$ {\it  Universit\'e Montpellier 2, Laboratoire Charles Coulomb,\\ UMR 5221,
F-34095 Montpellier, France}

\vspace{4mm}

${}^b$ {\it Laboratory of Theoretical Physics, JINR, 141980 Dubna, Moscow reg., Russia,\\
Moscow Institute of Physics and Technology, 141700, Dolgoprudny, Moscow reg., Russia,\\
Institute of Theoretical and Experimental Physics, 117259 Moscow, Russia}

\vspace{4mm}

${}^c$ {\it Laboratoire de Physique Th\'eorique LAPTH, CNRS and Universit\'e de Savoie,\\
BP 110, 74941 Annecy-le-Vieux Cedex, France}

\vspace{4mm}

${}^d$ {\it Steklov Mathematical Institute,
Moscow, Russia}

\vspace{4mm}

%{\bf \red{\today}}

\end{center}

\vspace{2mm}

\begin{abstract}
We study $GL(3)$-invariant integrable models solvable by the nested algebraic Bethe ansatz.
Different formulas are given for the Bethe vectors and the actions of the generators of the Yangian $\cY(\mathfrak{gl}_3)$
on Bethe vectors are considered. These actions are relevant for the calculation of correlation functions and form factors of local operators of the underlying quantum models.
\end{abstract}

\vspace{2mm}

%%%%%%%%%%%%%%%%%%%%%%%%%%%%%%%%%%%%%%%%%%%%%

\renewcommand*{\thefootnote}{\arabic{footnote}}
\addtocounter{footnote}{-1}
%%%%%%%%%%%%%%%%%%%%%%%%%%%%%%%%%%%

\section{Introduction}

The formulation of the quantum inverse scattering method, or algebraic Bethe ansatz (ABA),  by the Leningrad school \cite{FadSklT79} provides the eigenvectors and eigenvalues of the transfer matrix. The latter is the  generating function of the conserved quantities of a large family of quantum integrable models. The algebraic structures underlying this method are given by the quantum groups  that correspond to deformation of some Lie algebras \cite{Jim85,Dri85}. The transfer matrix eigenvectors are constructed from the representation theory of the quantum group. In order to construct these eigenvectors one first should prepare  \textbf{Bethe vectors (BV)}, depending
on a set of complex variables. At this point, some explanation on our terminology may be worth giving. When the aforementioned variables are generic complex numbers, these BV (sometimes called off-shell BV in the literature) are ``just'' vectors of the representation space, with no specific properties with respect to the transfer matrix.
When these variables satisfy a special set of coupled equations (Bethe equations), then the corresponding BV becomes an eigenvector of the transfer matrix and is called an
\textbf{on-shell Bethe vector}. In other words, on-shell BV are just a particular case of BV, for which the variables obey the  Bethe equations. Note also that the name BV is prescribed for vectors that become transfer matrix eigenvectors when the parameters obey Bethe equations. Thus, BV are not just ``any'' vector of the representation.
BV play an important role
in the study of form factors and correlation functions of local operators of the underlying quantum models.

\medskip

In this paper we will consider one of the simplest models that can be solved by the ABA, the $GL(N)$-invariant models. For these models, BV are constructed from the representation theory of the Yangian $\cY(\mathfrak{gl}_N)$,  where $\mathfrak{gl}_N$ is the Lie algebra associated to the group $GL(N)$. In the $N=2$ case, these models correspond to the XXX Heisenberg spin chain and their solutions, obtained from the ABA, were given in \cite{FadTak84}.
In many physically interesting cases, one should consider models with a higher rank symmetry
(see e.g. \cite{Sut68,Uli70,Sut75,PerS81}).
The first formulation of BV for $GL(N)$-invariant models was given by P.~P. Kulish and N.~Yu. Reshetikhin in \cite{KulRes83} where the nested algebraic Bethe ansatz (NABA) was introduced. These vectors are given by recursion on the rank of the algebra and use of the embedding $\cY(\mathfrak{gl}_{N-1}) \subset \cY(\mathfrak{gl}_N)$. This formulation is convenient for construction  eigenvectors of the transfer matrix since the commutation relations can be formulated with auxiliary space formalism and have a similar form to the original  $\cY(\mathfrak{gl}_{2})$ case.
Later on, a trace formula for BV was introduced by V. O. Tarasov and N. A. Varchenko \cite{TarVar93} who gave the BV in terms of the $R$-matrix and the monodromy matrix $T(u)$ of $\cY(\mathfrak{gl}_N)$.
This formulation was used to construct a solution of the quantum q-KZ equation from Jackson integrals. The case of $\mathfrak{gl}(m|n)$ superalgebras was performed in \cite{BelRag08}.

In principle, these approaches can be used to find explicit representations for BV in terms of the quantum algebra generators.
However, these formulas have not been obtained,
except for the trivial $\mathfrak{gl}_2$ case. In the case of the quantum algebra $\cU_q(\widehat {\mathfrak{gl}}_N)$ the BV can be formulated   using a certain projection of Drinfel'd currents  \cite{KhoPak05}. The equivalence with the trace formula of BV was shown in \cite{KhoPakT07,KhoPak08}. This formalism allows one to calculate the BV as an ordering of the currents \cite{FraKhoPR08,OPS10} and eventually to obtain explicit formulas for BV in terms of the generators and to get an integral presentation of their scalar products \cite{BelPakR10}. One of the goals of this
paper is to find an explicit  representation for the BV in the case of the Yangian $\cY(\mathfrak{gl}_3)$ starting
directly from the Yangian algebra.  We give  explicit and iteration formulas and we connect them to the trace formula  \cite{TarVar93,TarVar98}.

Another goal of this paper is to derive the action of the monodromy matrix entries on the BV. This action is very important
in the problem of the calculation of form factors and correlation functions. For a wide class of quantum integrable models
local operators can be expressed in terms of the monodromy matrix due to the inverse scattering problem \cite{KitMaiT99,MaiTer00}.
Then the calculation of their form factors and correlation functions can be reduced to the calculation of scalar
products of BV, provided that the action of the generators on the BV  gives a finite linear combination of BV.
The latter property is almost evident for $\cY(\mathfrak{gl}_2)$-based models; however it becomes very non-trivial
in the $\cY(\mathfrak{gl}_3)$ case.
In this paper we show that the action of the monodromy matrix entries on the BV does give a linear combination of BV. This
opens a way to study form factors of local operators via the scalar product formula obtained in \cite{Res86}.

The paper is organized as follows. Section \ref{S-DefNot} is devoted to definitions and notations.
Section~\ref{Yang} deals with the Yangian $\cY(\mathfrak{gl}_3)$ and its highest weight  representations (used in the construction of BV). We also give in this section some exchange relations between products of $T_{ij}(u)$ that, to the best of our knowledge were not previously known.
Section \ref{sec:BV} contains different formulas for the BV:  we give different iteration  and  explicit formulas  from ordered product of $\cY(\mathfrak{gl}_3)$ generators and show in section \ref{prf:BV}
that they are equivalent to the trace formula introduced in \cite{TarVar93,TarVar98}.
In  section \ref{sec:TijBV} we give the actions of the $T_{ij}(w)$ on the BV in terms of linear combinations of BV.
Section \ref{sec:proofs} and appendices contain the proofs of our different statements.

%%%%%%%%%%%%%%%%%%%%%%%%%%%%%%%%%%%%%%%%%%%%%

\section{Definitions and notations\label{S-DefNot}}

We consider the normalized  $R$-matrix
 \beq{R-mat}
 R(x,y)=\frac{\mathbf{I}+g(x,y)\mathbf{P}}{f(x,y)},
 \eeq
 with\footnote{$e_{ij}$ are the $3\times3$ elementary matrices.} $\mathbf{P}=\sum_{i=1}^3\sum_{j=1}^3 e_{ij}\otimes e_{ji}$,  and
 \beq{gf}
 g(x,y)=\frac{c}{x-y},\qquad
 f(x,y)=1+g(x,y)=\frac{x-y+c}{x-y},
\eeq
where $c$ is a constant. This $R$-matrix is $GL(3)$-invariant,
\beq{GL3inv}g_1g_2\,R(x,y)\,g_1^{-1}g_2^{-1}=R(x,y), \quad\forall g\in GL(3),
\eeq
which implies that it is also $\mathfrak{gl}_3$-invariant $[M_1+M_2\,,\,R(x,y)]=0$, $\forall M\in\mathfrak{gl}_3$.

The $R$-matrix \eqref{R-mat} satisfies Yang--Baxter equation
\beq{Y-B}
R_{12}(x,y)R_{13}(x,z)R_{23}(y,z)=R_{23}(y,z)R_{13}(x,z)R_{12}(x,y).
\eeq
 Equation \eqref{Y-B} holds in the tensor product
$\mathbb{C}^3\otimes\mathbb{C}^3\otimes\mathbb{C}^3$.
The $R$-matrix subscripts  in \eqref{Y-B} show the spaces where it acts
non-trivially.

Apart from the functions $g(x,y)$ and $f(x,y)$ we also introduce the functions
\ben\label{univ-not}
 h(x,y)=\frac{f(x,y)}{g(x,y)}=\frac{x-y+c}{c}
 \mb{and} t(x,y)=\frac{g(x,y)}{h(x,y)}=\frac{c^2}{(x-y+c)(x-y)}.
\een

We always denote sets of variables by bar: $\bar w$, $\bar u$, $\bar v$ etc.
Individual elements
of the sets are denoted by subscripts and without bar: $w_j$, $u_k$, $v_{\ell}$ etc. As a rule, the number of elements in the
sets is not shown explicitly; however we give these cardinalities in
special comments after the formulas.
Subsets of variables are denoted by Roman subscripts: $\bar u_{\so}$, $\bar v_{\st}$, $\bar w_{\rm ii}$
etc. For example, the notation $\bar u\Rightarrow\{\bar u_{\so},\;\bar u_{\st}\}$ means that the
set $\bar u$ is divided into two disjoint subsets $\bar u_{\so}$ and $\bar u_{\st}$.
We assume that the elements in every subset are ordered in such a way that the sequence of
their subscripts is strictly increasing.
For the union of two sets into another one we use the notation $\{ \bar u, \bar w\}=\bar \xi$.
Finally we use a  special notation $\bar u_{j}$, $\bar v_{k}$ and so on  for the sets
$\bar u\setminus u_j$, $\bar v\setminus v_k$ etc.

In order to avoid excessively cumbersome formulas we use shorthand notation for products of
functions depending on one or two variables. Namely, whenever such a function depends on a set of variables, this means that we
deal with the product of this function with respect to the corresponding set, as follows:
 \beq{SH-prod}
 \lambda_i(\bar u)=\prod_{u_j\in\bar u}  \lambda_i(u_j);\quad
 g(x_k, \bar w_\ell)= \prod_{\substack{w_j\in\bar w\\w_j\ne w_\ell}} g(x_k, w_j);\quad
 f(\bar u_{\st},\bar u_{\so})=\prod_{u_j\in\bar u_{\st}}\prod_{u_k\in\bar u_{\so}} f(u_j,u_k).
 \eeq
This notation is also used for the product of commuting operators,
\ben\label{SH-prod-O}
T_{ij}(\bar u)=\prod_{u_k\in\bar u} T_{ij}(u_k).
\een

In various formulas the Izergin--Korepin determinant $\Izer_k(\bar x|\bar y)$ appears\footnote{
Note that by definition this function  depends on two sets of variables. Therefore, the convention on shorthand notations for
the products is not applicable in this case.}.
It is defined for two sets $\bar x$ and $\bar y$ with common cardinality $\# \bar x=\# \bar y=k$,
 \begin{equation}\label{K-def}
 \Izer_k(\bar x|\bar y)=\prod_{\ell<m}^k g(x_\ell,x_m)g(y_m,y_\ell)\cdot h(\bar x,\bar y)
 \; \det_k\left[ t(x_i,y_j) \right].
\end{equation}

%%%%%%%%%%%%%%%%%%%%%%%%%%%%%%%%%%%%%%%%%%%%%

\section{$\cY(\mathfrak{gl}_3)$ Yangian  and its highest weights representation \label{Yang}}
The $R$-matrix (\ref{R-mat}) allows one to formulate the Yangian commutation relations as a single equality. For such a purpose, the generators are gathered in a (matrix valued)  $\LL$-operator
$$\LL(z)=\sum_{i,j=1}^3 e_{ij}\otimes \LL_{ij}(z)\in\mbox{End}(\CC^3)\otimes\cY(\mathfrak{gl}_3)[[z^{-1}]]\,,$$
where the $3\times3$ matrices $e_{ij}$ span a so-called auxiliary space.
The Yangian
 commutation relation are then expressed as a single relation
\ben\label{RTT}
R_{12}(u,v)\LL_1(u)\LL_2(v)=\LL_2(v)\LL_1(u)R_{12}(u,v),
\een
where we used the notation
$\LL_{k}(z)\in \sk{\CC^3}^{\otimes M}\ot \Yan$ for the 
$\LL$-operator acting nontrivially on the $k$-th
tensor factor in the product $\sk{\CC^{3}}^{\otimes M}$ for $1\leq k\leq M$.

From the RTT presentation (\ref{RTT}) it is clear (see e.g. \cite{MNO}) that
\beq{def-phi}
\vph\,:\,\begin{cases} \cY(\mathfrak{gl}_N)\quad \to\quad \cY(\mathfrak{gl}_N)\\
T(u) \quad\mapsto\quad T^t(-u)\end{cases}
\eeq
defines an isomorphism of the Yangian for any choice of transposition $^t$. Here, we choose as the transposition
\beq{transpo}
e_{ij}^t = e_{N+1-j,N+1-i}\,.
\eeq
Obviously, the isomorphism $\vph$ is idempotent, $\vph^2=id$.

From \eqref{RTT} we extract  the following commutation relations:
\ben
[\LL_{ij}(u),\LL_{kl}(v)] &=& g(u,v)\Big(\LL_{kj}(v)\,\LL_{il}(u)-\LL_{kj}(u)\,\LL_{il}(v)\Big)
\label{Ycom}\\
&=& g(u,v)\Big(\LL_{il}(u)\,\LL_{kj}(v)-\LL_{il}(v)\,\LL_{kj}(u)\Big).\label{Ycom2}
\een
These commutation relations imply in particular for multiple products $T_{ij}(\bar z)=T_{ij}(z_1)...T_{ij}(z_m)$
\ben
\label{TT-triv}
T_{ij}(\bar y)T_{ij}(\bar x)&=& T_{ij}(\bar x)T_{ij}(\bar y),\quad 1\leq i, j\leq 3.
\een
They also lead to the following property:
\begin{prop}\label{Lem1}
In $\cY(\mathfrak{gl}_3)$, we have the following multiple exchange relations, written for arbitrary $i$, $j$ and $k$:
\beq{two}
T_{ij}(\bar y)T_{ik}(\bar x)=(-1)^{\nx}\sum K_{\nx}(\bar x|\bar w_{\st}+c)f(\bar w_{\st},\bar w_{\so})
T_{ik}(\bar w_{\st})T_{ij}(\bar w_{\so}).
\eeq
\beq{one}
T_{ij}(\bar y)T_{kj}(\bar x)=(-1)^{\nx}\sum K_{\nx}(\bar w_{\st}|\bar x+c)f(\bar w_{\so},\bar w_{\st})
T_{kj}(\bar w_{\st})T_{ij}(\bar w_{\so}),
\eeq
Here $\{\bar x,\bar y\}=\bar w$, $\#\bar x=\nx$, $\#\bar y=n_y$ and the sums are taken
over  partitions $\bar w\Rightarrow\{\bar w_{\so},\bar w_{\st}\}$ with $\#\bar w_{\st}=n_x$ and $\#\bar w_{\so}=n_y$.
$\Izer_{\nx}(\bar x|\bar y)$ is the Izergin--Korepin determinant \eqref{K-def}.

These formulas have twins
\beq{two-t}
T_{ij}(\bar y)T_{ik}(\bar x)=(-1)^{n_y}\sum K_{n_y}(\bar  w_{\so}|\bar y+c)f(\bar w_{\st},\bar w_{\so})
T_{ik}(\bar w_{\st})T_{ij}(\bar w_{\so}).
\eeq
\beq{one-t}
T_{ij}(\bar y)T_{kj}(\bar x)=(-1)^{n_y}\sum K_{n_y}(\bar y|\bar w_{\so}+c)f(\bar w_{\so},\bar w_{\st})
T_{kj}(\bar w_{\st})T_{ij}(\bar w_{\so}),
\eeq
\end{prop}

The twin formulas follow from \eqref{two} and \eqref{one} due to different reductions \eqref{K-K}
of the Izergin--Korepin determinants $K_{\nx+n_y}(\bar w|\{\bar w_{\so},\bar x+c\})$ and $K_{\nx+n_y}(\{\bar x, \bar w_{\so}\}|\bar w+c)$ respectively, as follows:
\beq{Red-BigK}
\begin{array}{l}
K_{\nx+n_y}(\bar w|\{\bar w_{\so}+c,\bar x+c\})=(-1)^{n_y} K_{\nx}(\bar w_{\st}|\bar x+c)= (-1)^{\nx}K_{n_y}(\bar y|\bar w_{\so}+c),\\
K_{\nx+n_y}(\{\bar x, \bar w_{\so}\}|\bar w+c)=(-1)^{n_y}  K_{\nx}(\bar x|\bar w_{\st}+c)=(-1)^{\nx}K_{n_y}(\bar  w_{\so}|\bar y+c).
\end{array}
\eeq
The commutation relations \eqref{two} and \eqref{one} are related by the morphism $\varphi$. The
proof of \eqref{two} is given in appendix \ref{prf:TT}.
 \paragraph{Remark:} When all indices $i,j,k$ are equal, the l.h.s. of \eqref{two}, \eqref{one} look rather different from their r.h.s. However, it can be proven through recursion that indeed $\sum K_{\nx}(\bar x|\bar w_{\st}+c)f(\bar w_{\so},\bar w_{\st})=(-1)^{n_x}$ (and similarly for \eqref{one}), so that l.h.s. and r.h.s. coincide.

\medskip

\begin{cor}\label{mult-sl2}
The multiple exchange relations \eqref{two} and \eqref{one} are also valid in $\cY(\mathfrak{gl}_2)$,
for $1\leq i ,j, k\leq 2$.
They provide the multiple exchange relations between the Cartan-like generators $A(x)=T_{11}(x)$,
 $D(x)=T_{22}(x)$, and the generators $B(x)=T_{12}(x)$, $C(x)=T_{21}(x)$.
\end{cor}

To the best of our knowledge this compact form of multiple commutation relations in the $\cY(\mathfrak{gl}_2)$ case was not known
previously (see, however, \cite{KitMaiT00,KitMST05} for various forms of the multiple action).

\paragraph{Example} To illustrate the formulas above, we consider commutation relations of the single
operator $T_{11}$ with the products of the operators $T_{12}$ and $T_{21}$.
Let us fix $i=j=1$,  $k=2$ in \eqref{two-t}. We also set $n_y=\#\bar y=1$, while $n_x=\#\bar x$ remains arbitrary. Then we
have
\begin{equation}\label{ex1}
T_{11}(y)T_{12}(\bar x) =  f(\bar x,y)\, T_{12}(\bar x)T_{11}(y)-\sum_{\ell=1}^{n_x} g(x_{\ell},y) f(\bar x_{\ell},x_{\ell})
\, T_{12}(\{\bar x_{\ell},y\})\, T_{11}(x_{\ell}).
\end{equation}
On the other hand, setting in \eqref{one} $i=2$,  $j=k=1$,
$n_x=1$ and keeping $n_y$ free we obtain
\begin{equation}\label{ex2}
T_{21}(\bar y) T_{11}( x) =  f( x, \bar y)\, T_{11}(x)T_{21}(\bar y)-\sum_{\ell=1}^{n_y} g(x,y_{\ell}) f( y_{\ell},\bar y_{\ell})
\, T_{11}(y_{\ell})\, T_{21}(\{\bar y_{\ell},x\}).
\end{equation}
One can easily recognize in these equations the standard formulas of the ABA.
\medskip

\section{Bethe vectors\label{sec:BV}}

We will construct the BV in finite dimensional highest weight representations of the Yangian $\Yan$. A highest weight  representation is freely generated by
a right highest weight vector $\rvec$, itself defined by:
\begin{equation}\label{hwv}
\LL_{ij}(u)\ \rvec=0\,,\quad 1\leq j<i\leq3\,,\quad \LL_{ii}(u)\rvec =\as_i(u)\ \rvec\,,\quad
i=1,2,3\,.
\end{equation}
Here $\as_i(u)$, $i=1,2,3$ are the weights of the representation.

Note that if $\cV\big(\lambda_1(u),...\lambda_j(u),...,\lambda_N(u)\big)$ is a representation of $\cY(\mathfrak{gl}_N)$, then, the
morphism $\vph$ maps it to the representation $\cV\big(\lambda_N(-u),...\lambda_{N+1-j}(-u),...,\lambda_1(-u)\big)$. Thus $\vph$ induces a map between BV constructed in different highest weight representations.

We denote the BV as $\bbb^{a,b}(\bar u;\bar v)$. They depend on two sets of parameters $\bar u$ and $\bar v$ with
$\#\bar u=a$ and $\#\bar v=b$.  As mentioned in the introduction, for BV, these parameters are generic complex numbers. We will see in section \ref{sec:on-shell} that when these parameters obey the Bethe equations, $\bbb^{a,b}(\bar u;\bar v)$ becomes a transfer matrix eigenvector, deserving the name on-shell BV for such a vector.

\subsection{Explicit formulas\label{sect:expl}}

An explicit formulation in terms of the generators $T_{12}(u)$, $T_{23}(v)$ and $T_{13}(x)$ can be written for BV.   These
representations involve summation over
partitions of the sets $\bar u$ and $\bar v$:
\begin{align}
\label{BV-explicit-1}
&&\bbb^{a,b}(\bar u;\bar v)=
\sum
\frac{\Izer_k(\bar v_{\so}|\bar u_{\so})}{ \as_2(\bar v_{\st})\as_2(\bar u)}\,
\frac{f(\bar v_{\st},\bar v_{\so})f(\bar u_{\st},\bar u_{\so})}
{f(\bar v_{\st},\bar u)f(\bar v_{\so},\bar u_{\so})}
\, \LL_{12}(\bar u_{\st})\LL_{13}(\bar u_{\so}) \LL_{23}(\bar v_{\st})\rvec,
\\
\label{BV-explicit-2}
&&\bbb^{a,b}(\bar u;\bar v)=
\sum
\frac{\Izer_k(\bar v_{\so}| \bar u_{\so})}{ \as_2(\bar u_{\st})\as_2(\bar v)}\,
\frac{ f(\bar v_{\so},\bar v_{\st})f(\bar u_{\so},\bar u_{\st})}
{f(\bar v_{\so},\bar u_{\so})f(\bar v,\bar u_{\st})}\,  \LL_{23}(\bar v_{\st})\LL_{13}(\bar v_{\so}) \LL_{12}(\bar u_{\st})\rvec,
\\
\label{Or-form}
&&\bbb^{a,b}(\bla;\bmu) =\sum \frac{\Izer_{k}(\bmu_{\so}|\bla_{\so})}{\as_2(\bar v_{\st})\as_2(\bar u)}
\frac{f(\bmu_{\st},\bmu_{\so})f(\bla_{\so},\bla_{\st})}{f(\bmu,\bla)}\,
T_{13}(\bla_{\so})T_{12}(\bla_{\st})T_{23}(\bmu_{\st})|0\rangle,
\\
\label{Or-form2}
&&\bbb^{a,b}(\bla;\bmu) =\sum \frac{\Izer_{k}({\bmu}_{\so}|{\bla}_{\so})}{\as_2(\bar u_{\st})\as_2(\bar v)}
\frac{f(\bmu_{\st},\bmu_{\so})f(\bla_{\so},\bla_{\st})}{f(\bmu,\bla)}\,
T_{13}({\bmu}_{\so})T_{23}({\bmu}_{\st})T_{12}({\bla_{\st}})|0\rangle.
\end{align}

Here the sums are taken over partitions of the sets $\bar u\Rightarrow\{\bar u_{\so},\bar u_{\st}\}$ and $\bar v\Rightarrow\{\bar v_{\so}$, $\bar v_{\st}\}$ with $0\leq\#\bar u_{\so}=\#\bar v_{\so}\leq\mbox{min}(a,b)$. We used the notation
$k=\#\bar u_{\so}=\#\bar v_{\so}$. $ \Izer_k(\bar v_{\so}|\bar u_{\so})$ is the Izergin--Korepin determinant \eqref{K-def}.

 These representations can be derived in the framework of the current approach to the NABA (see e.g. \cite{OPS10}).
However we do not use this method. Instead we show directly that the vectors defined above are Bethe vectors (that is, that they become eigenvectors of the transfer matrix, provided the parameters $\bar u$ and $\bar v$ satisfy Bethe equations.)

In section \ref{prf:BV}, we show that these explicit formulas are all equivalent.
One can already remark, however, that
\eqref{BV-explicit-1} and \eqref{BV-explicit-2} (as well as \eqref{Or-form} and \eqref{Or-form2}) are related by the morphism $\vph$, provided
\beq{phiBV}
\vph \big(\bbb^{a,b}(\bar u;\bar v)\big) = \bbb^{b,a}(-\bar v;-\bar u),
\eeq
where $\bbb^{a,b}(\bar u;\bar v)$ is constructed in the representation
$\cV\big(\lambda_1(u),\lambda_2(u),\lambda_3(u)\big)$, while $\bbb^{b,a}(-\bar v;-\bar u)$ lies in the
representation $\cV\big(\lambda_3(-u),\lambda_2(-u),\lambda_1(-u)\big)$.  We show the action of
the morphism $\vph$ on the vectors \eqref{BV-explicit-1}--\eqref{Or-form2} explicitly in section \ref{prf:phi}.

\subsection{Iteration formulas\label{sect:iter}}

Here we give  iteration formulas that allow to build BV in a recursive way.
There are essentially two iteration formulas for  BV, depending on which set, $\bar u$ or $\bar v$, one wishes to make a recursion. The first one reads:
 \begin{equation}
 \label{iter1}
\lambda_2(u_k)f(\bar v,u_k)\bbb^{a,b}(\bar u;\bar v)=  \LL_{12}(u_k) \bbb^{a-1,b}(\bar u_k;\bar v)
+ \sum_{i=1}^b g( v_{i},u_k)f(\bar v_{i}, v_{i})  \LL_{13}(u_k)\bbb^{a-1,b-1}(\bar u_k; \bar v_i).
\end{equation}
Here $u_k$ is an arbitrary element from the set $\bar u$. The second recursion has the form
 \begin{equation}
 \label{iter2}
\lambda_2(v_{k})f(v_{k},\bar u) \bbb^{a,b}(\bar u;\bar v)=  \LL_{23}(v_{k}) \bbb^{a,b-1}(\bar u;\bar v_{k})
+ \sum_{j=1}^a g(v_{k},u_j)f(u_{j}, \bar u_{j}) \LL_{13}(v_{k})\bbb^{a-1,b-1}(\bar u_{j}; \bar v_{k}),
\end{equation}
where now $v_k$ is an arbitrary element from the set $\bar v$.
The initial conditions are given by
\ben\label{BV-initial}
&&\lambda_2(\bar \mu) \bbb^{0,b}(\bar\mu)=\,\LL_{23}(\bar \mu) \rvec\mb{and}\lambda_2(\bar \la) \bbb^{a,0}(\bar\la)=\,\LL_{12}(\bar \la) \rvec.
\een

These formulas can be  extracted from the explicit representations \eqref{BV-explicit-1}--\eqref{Or-form2}
(see section~\ref{sect:iter-explicit}). On the other hand
they can be obtained from the trace formula (\ref{trace1}) (see section~\ref{prf:iter}). Any of the above two recursions together with the initial condition \eqref{BV-initial} defines uniquely the vectors $\bbb^{a,b}(\bar u;\bar v)$. As it is already known that the trace formula defines the BV \cite{TarVar98}, it proves that the explicit representations \eqref{BV-explicit-1}--\eqref{Or-form2}
also give BV.

\section{Multiple action of the monodromy matrix on Bethe vectors\label{sec:TijBV}}

In various models the form factors of local operators can be reduced to matrix elements of the monodromy
matrix entries $T_{ij}$. In order to calculate such matrix elements, one first of all should evaluate the
actions of $T_{ij}$ on the BV.
In  models described by the Yangian $\cY(\mathfrak{gl}_{2})$ this action  evidently can be expressed as a linear combination of BV. One can ask whether the same effect takes place in the case of $\cY(\mathfrak{gl}_{3})$-based models. Indeed, the explicit
expressions for BV  \eqref{BV-explicit-1}--\eqref{Or-form2} are very specific polynomials in generators $T_{k\ell}$ (with $k<\ell$) acting on the highest weight vector $|0\rangle$. It is not obvious that the result of the action of $T_{ij}$ on such polynomials can be presented
as a finite linear combination of the polynomials of the same type. If such a representation is impossible, then the  form
factors of $T_{ij}$  can not be reduced to the scalar products of BV, as they were in the $\cY(\mathfrak{gl}_{2})$ case.
In this section we give the list of formulas, showing that any multiple action of the monodromy matrix entries on the BV is
a finite linear combination of BV.

Below everywhere $\{\bar\mu,\bar w\}=\bar\xi$, $\{\bla,\bar w\}=\bar\eta$ and $\#\bar w=n$.  We also use the notation
$$
\fr_j(w)=\frac{\lambda_j(w)}{\lambda_2(w)}\,,\quad j=1,3.
$$

\begin{itemize}

\item Multiple action of $T_{13}$
 \beq{act13}
 T_{13}(\bar w)\mathbb{B}^{a,b}(\bla;\bmu)=\lambda_2(\bar w)\,\mathbb{B}^{a+n,b+n}(\bar\eta;\bar\xi).
 \eeq

\item Multiple action of $T_{12}$
 \beq{act12}
 T_{12}(\bar w)\mathbb{B}^{a,b}(\bla;\bmu)=(-1)^n\lambda_2(\bar w)\,\sum
 f(\bar\xi_{\st},\bar\xi_{\so})\Izer_n(\bar\xi_{\so}|\bar w+c)\,
 \mathbb{B}^{a+n,b}(\bar\eta;\bar\xi_{\st}).
 \eeq
The sum is taken over partitions of $\bar\xi\Rightarrow\{\bar\xi_{\so},\bar\xi_{\st}\}$ with $\#\bar\xi_{\so}=n$.

\item Multiple action of $T_{23}$
 \beq{act23}
 T_{23}(\bar w)\mathbb{B}^{a,b}(\bla;\bmu)=(-1)^n\lambda_2(\bar w)\,\sum
 f(\bar\eta_{\so},\bar\eta_{\st})\Izer_n(\bar w|\bar\eta_{\so}+c)\,
 \mathbb{B}^{a,b+n}(\bar\eta_{\st};\bar\xi).
 \eeq
The sum is taken over partitions of $\bar\eta\Rightarrow\{\bar\eta_{\so},\bar\eta_{\st}\}$
 with $\#\bar\eta_{\so}=n$.

\item Multiple action of $T_{22}$
 \beq{act22}
 T_{22}(\bar w)\mathbb{B}^{a,b}(\bla;\bmu)=\lambda_2(\bar w)\,\sum f(\bar\xi_{\st},\bar\xi_{\so})
 f(\bar\eta_{\so},\bar\eta_{\st})\Izer_n(\bar\xi_{\so}|\bar w+c)
 \Izer_n(\bar w|\bar\eta_{\so}+c)\,\mathbb{B}^{a,b}(\bar\eta_{\st};\bar\xi_{\st}).
 \eeq
\begin{tabular}{ll}
 The sum is taken over partitions of:& $\bar\eta\Rightarrow\{\bar\eta_{\so},\bar\eta_{\st}\}$
 with $\#\bar\eta_{\so}=n$; \\
& $\bar\xi\Rightarrow\{\bar\xi_{\so},\bar\xi_{\st}\}$ with $\#\bar\xi_{\so}=n$.
\end{tabular}

\item Multiple action of $T_{11}$
 \beq{act11}
 T_{11}(\bar w)\mathbb{B}^{a,b}(\bla;\bmu)=\lambda_2(\bar w)\,\sum
\fr_1(\bar\eta_{\so})\, \frac{f(\bar\xi_{\st},\bar\xi_{\so})
 f(\bar\eta_{\st},\bar\eta_{\so})}{f(\bar\xi_{\st},\bar\eta_{\so})}   \Izer_n(\bar\xi_{\so}|\bar w+c)
 \Izer_n(\bar\eta_{\so}|\bar\xi_{\so}+c)\,\mathbb{B}^{a,b}(\bar\eta_{\st};\bar\xi_{\st}).
 \eeq
\begin{tabular}{ll}
 The sum is taken over partitions of:& $\bar\eta\Rightarrow\{\bar\eta_{\so},\bar\eta_{\st}\}$
 with $\#\bar\eta_{\so}=n$; \\
& $\bar\xi\Rightarrow\{\bar\xi_{\so},\bar\xi_{\st}\}$ with $\#\bar\xi_{\so}=n$.
\end{tabular}

\item Multiple action of $T_{33}$
 \beq{act33}
 T_{33}(\bar w)\mathbb{B}^{a,b}(\bla;\bmu)=\lambda_2(\bar w)\,\sum
\fr_3(\bar\xi_{\so})\,\frac{f(\bar\xi_{\so},\bar\xi_{\st})
 f(\bar\eta_{\so},\bar\eta_{\st})}{f(\bar\xi_{\so},\bar\eta_{\st})}   \Izer_n(\bar w|\bar\eta_{\so}+c)
 \Izer_n(\bar\eta_{\so}|\bar\xi_{\so}+c)\,\mathbb{B}^{a,b}(\bar\eta_{\st};\bar\xi_{\st}).
 \eeq
\begin{tabular}{ll}
 The sum is taken over partitions of:& $\bar\eta\Rightarrow\{\bar\eta_{\so},\bar\eta_{\st}\}$
 with $\#\bar\eta_{\so}=n$; \\
& $\bar\xi\Rightarrow\{\bar\xi_{\so},\bar\xi_{\st}\}$ with $\#\bar\xi_{\so}=n$.
\end{tabular}

\item Multiple action of $T_{21}$
 \begin{multline}\label{act21}
 T_{21}(\bar w)\mathbb{B}^{a,b}(\bla;\bmu)=(-1)^n\lambda_2(\bar w)\,\sum
\fr_1(\bar\eta_{\so})\,
f(\bar\eta_{\st},\bar\eta_{\so})f(\bar\eta_{\st},\bar\eta_{\sth})f(\bar\eta_{\sth},\bar\eta_{\so})
 \frac{f(\bar\xi_{\st},\bar\xi_{\so})}{f(\bar\xi_{\st},\bar\eta_{\so})} \\
 \times  \Izer_n(\bar w|\bar\eta_{\st}+c)
  \Izer_n(\bar\eta_{\so}|\bar\xi_{\so}+c)\Izer_n(\bar\xi_{\so}|\bar w+c)
 \,\mathbb{B}^{a-n,b}(\bar\eta_{\sth};\bar\xi_{\st}).
 \end{multline}
 \begin{tabular}{ll}
The sum is taken over partitions of: & $\bar\eta\Rightarrow\{\bar\eta_{\so},\bar\eta_{\st},\bar\eta_{\sth}\}$ with $\#\bar\eta_{\so}=\#\bar\eta_{\st}=n$; \\
& $\bar\xi\Rightarrow\{\bar\xi_{\so},\bar\xi_{\st}\}$ with $\#\bar\xi_{\so}=n$.
\end{tabular}

\item Multiple action of $T_{32}$
 \begin{multline}\label{act32}
 T_{32}(\bar w)\mathbb{B}^{a,b}(\bla;\bmu)=
 (-1)^n\lambda_2(\bar w)\,\sum \fr_3(\bar\xi_{\so})\,
f(\bar\xi_{\so},\bar\xi_{\st})f(\bar\xi_{\so},\bar\xi_{\sth})f(\bar\xi_{\sth},\bar\xi_{\st})
 \frac{f(\bar\eta_{\so},\bar\eta_{\st})}{f(\bar\xi_{\so},\bar\eta_{\st})} \\
 \times  \Izer_n(\bar w|\bar\eta_{\so}+c)
 \Izer_n(\bar\eta_{\so}|\bar\xi_{\so}+c)\Izer_n(\bar\xi_{\st}|\bar w+c)
 \,\mathbb{B}^{a,b-n}(\bar\eta_{\st};\bar\xi_{\sth}).
 \end{multline}
 \begin{tabular}{ll}
The sum is taken over partitions of: & $\bar\xi\Rightarrow\{\bar\xi_{\so},\bar\xi_{\st},\bar\xi_{\sth}\}$ with $\#\bar\xi_{\so}=\#\bar\xi_{\st}=n$; \\
&  $\bar\eta\Rightarrow\{\bar\eta_{\so},\bar\eta_{\st}\}$
 with $\#\bar\eta_{\so}=n$.
\end{tabular}

\item Multiple action of $T_{31}$
 \begin{multline}\label{act31}
 \hspace{-5mm}T_{31}(\bar w)\mathbb{B}^{a,b}(\bla;\bmu)=\lambda_2(\bar w)\,\sum
\fr_1(\bar\eta_{\st})\,\fr_3(\bar\xi_{\so})\,
 \Izer_n(\bar\eta_{\so}|\bar\xi_{\so}+c)\Izer_n(\bar\eta_{\st}|\bar\xi_{\st}+c)\Izer_n(\bar w|\bar\eta_{\so}+c)
 \Izer_n(\bar\xi_{\st}|\bar w+c) \\
 \times
 \frac{f(\bar\eta_{\so},\bar\eta_{\st})f(\bar\eta_{\so},\bar\eta_{\sth})f(\bar\eta_{\sth},\bar\eta_{\st})
 f(\bar\xi_{\so},\bar\xi_{\st})f(\bar\xi_{\so},\bar\xi_{\sth})f(\bar\xi_{\sth},\bar\xi_{\st})}
 {f(\bar\xi_{\so},\bar\eta_{\st})f(\bar\xi_{\so},\bar\eta_{\sth})f(\bar\xi_{\sth},\bar\eta_{\st})}\,
 \mathbb{B}^{a-n,b-n}(\bar\eta_{\sth};\bar\xi_{\sth}).
 \end{multline}

 \begin{tabular}{ll}
The sum is taken over partitions of:&
 $\bar\xi\Rightarrow\{\bar\xi_{\so},\bar\xi_{\st},\bar\xi_{\sth}\}$ with $\#\bar\xi_{\so}=\#\bar\xi_{\st}=n$; \\
& $\bar\eta\Rightarrow\{\bar\eta_{\so},\bar\eta_{\st},\bar\eta_{\sth}\}$ with $\#\bar\eta_{\so}=\#\bar\eta_{\st}=n$.
\end{tabular}
\end{itemize}
Proofs are given in sections~\ref{proofT13}, \ref{proofT12}, \ref{proofT22}.

As we have explained at the beginning of this section, the above formulas for the multiple action allow one to reduce the calculation
of the form factors of $T_{ij}$ to the calculation of scalar products of BV. This means that one can use the results of
\cite{Res86}, where a representation for the general case of a scalar product of two BV was obtained. Although
this general formula is quite cumbersome, our recent results \cite{BelPakRS12c,BelPakRS12d} show that it may lead to reasonable representations for form factors.

\subsection{On-shell Bethe vectors and Bethe equations\label{sec:on-shell}}

The formulas for the multiple action \eqref{act22}--\eqref{act33} provide us with a simple proof that the explicit expressions given in section \ref{sect:expl} indeed correspond to BV. For this it is enough to show that if the sets of parameters $\bar u$ and $\bar v$ satisfy Bethe equations, then these vectors become eigenvectors of the transfer matrix
 \beq{transfert}
 t(w)=\mbox{Tr}\big(\LL(w)\big)=\LL_{11}(w)+\LL_{22}(w)+\LL_{33}(w).
 \eeq
Let us check this. Observe that in the case $n=1$ the above multiple actions imply
\ben\label{actA}
 T_{11}( w)\mathbb{B}^{a,b}(\bla;\bmu)&=&
 -\lambda_2( w)\,\sum
 \fr_1(\bar\eta_{\so})
 \frac{f(\bar\xi_{\st},\bar\xi_{\so}) f(\bar\eta_{\st},\bar\eta_{\so})}{f(\bar\xi_{\st},\bar\eta_{\so})}
 \frac{ g(\bar\xi_{\so},w+c)}{h(\bar\xi_{\so},\bar\eta_{\so})}\,\mathbb{B}^{a,b}(\bar\eta_{\st};\bar\xi_{\st}),
\quad\qquad \\
 \label{actD1}
 T_{22}( w)\mathbb{B}^{a,b}(\bla;\bmu)&=&
 \lambda_2( w)\,\sum f(\bar\xi_{\st},\bar\xi_{\so})
 f(\bar\eta_{\so},\bar\eta_{\st})g(\bar\xi_{\so}, w+c)
g( w,\bar\eta_{\so}+c)\,\mathbb{B}^{a,b}(\bar\eta_{\st};\bar\xi_{\st}),
 \\
\label{actD2}
 T_{33}( w)\mathbb{B}^{a,b}(\bla;\bmu)&=&
 -\lambda_2( w)\,\sum  \fr_3(\bar\xi_{\so})\,
 \frac{f(\bar\xi_{\so},\bar\xi_{\st})f(\bar\eta_{\so},\bar\eta_{\st})}{f(\bar\xi_{\so},\bar\eta_{\st})}
 \frac{g( w,\bar\eta_{\so}+c)}{h(\bar\xi_{\so},\bar\eta_{\so})}\,
 \mathbb{B}^{a,b}(\bar\eta_{\st};\bar\xi_{\st}),
 \een
where now sums are taken over partitions of $\{\bar\mu, w\}=\bar\xi\Rightarrow\{\bar\xi_{\so}, \bar\xi_{\st}\}$ and
$\{\bla, w\}=\bar\eta\Rightarrow\{\bar\eta_{\so}, \bar\eta_{\st}\}$ with $\#\bar\xi_{\so}=\#\bar\eta_{\so}=1$.
Making these sums explicit, we get the action of the transfer matrix \eqref{transfert} on a BV:
\ben
&&t(w)\,\mathbb{B}^{a,b}(\bar u;\bmu) =
\Big(\lambda_1(w)f(\bar u,w)+\lambda_2(w)f(w,\bar u)f(\bar v,w)+\lambda_3(w)f(w,\bar v)\Big)
\,\mathbb{B}^{a,b}(\bar u;\bmu)
\nonu
&&\qquad
+\lambda_2(w)\,f(\bar v,w)\sum_{j=1}^a g(w,u_j)\Big( \fr_1(u_j) \frac{f(\bar u_j,u_j)}{f(\bar v, u_j)}-f(u_j,\bar u_j)
\Big)\, \mathbb{B}^{a,b}(\{\bar u_j,w\};\bar v)
\nonu
&&\qquad
+\lambda_2(w)\,f(w,\bar u)\sum_{i=1}^b g(w,v_i)\Big(\fr_3(v_i) \frac{f(v_i,\bar v_i)}{f(v_i,\bar u)}-
f(\bar v_i, v_i)
\Big)\, \mathbb{B}^{a,b}(\bar u;\{\bar v_i,w\})
\label{t-B-again}\\
&&\qquad
+\lambda_2(w)\sum_{i=1}^b\sum_{j=1}^a g(u_j,v_i)\,\left\{ g(w,v_i)f(\bar v_i,v_i)\Big(\fr_1(u_j) \frac{f(\bar u_j,u_j)}{f(\bar v, u_j)}-
f(u_j,\bar u_j)
\Big)\right.
\nonu
&&\qquad\qquad\qquad\left.
+g(u_j,w)f(u_j,\bar u_j)\Big(\fr_3(v_i) \frac{f(v_i,\bar v_i)}{f(v_i,\bar u)}-
f(\bar v_i, v_i)
\Big)\,
\right\} \mathbb{B}^{a,b}(\{\bar u_j,w\};\{\bar v_i,w\}).
\qquad
\nonumber
\een
To obtain the  two last lines, we have used
$g(w,v_{i})g(u_{j},w)=g(u_{j},v_{i})\big(g(w,v_{i})+g(u_{j},w)\big)$.

Demanding that each term of the second and third line in \eqref{t-B-again} identically vanishes,
one recovers the Bethe equations for the Bethe roots $\{\bar u; \bar v\}$:
\ben\label{Beqs1}
\fr_1(u_i)f(\bar u_i,u_i) &=& f(u_i,\bar u_i)f(\bar v, u_i), \\
\fr_3(v_i)f(v_i,\bar v_i) &=& f(v_i,\bar u)f(\bar v_i, v_i).\label{Beqs2}
\een
It implies also that the last two lines vanish.
This shows that $\mathbb{B}^{a,b}(\bla;\bmu)$ is an eigenvector of $t(w)$, provided the sets $\bar u$ and $\bar v$
satisfy the equations \eqref{Beqs1}, \eqref{Beqs2}. Hence, $\mathbb{B}^{a,b}(\bla;\bmu)$ for generic complex
$\bar u$ and $\bar v$ is a BV. The eigenvalue $\Lambda^{a,b}(w; \bar\la; \bar\mu)$ of an on-shell BV is given by the first line of
\eqref{t-B-again}:
 \ben\label{Bvp}
 \Lambda^{a,b}(w; \bar\la; \bar\mu)=
 \lambda_1(w)f(\bar u,w)+\lambda_2(w)f(w,\bar u)f(\bar v,w)+\lambda_3(w)f(w,\bar v).
\een

\subsection{Trace formula}

For completeness we recall the trace formula given in \cite{TarVar93,TarVar98}.
Consider  the set of  variables
$(\bar\la; \bar\muu) = (
\la_{1},\ldots,\la_{a}; \muu_{1},\ldots,
\muu_{b})$, the set of auxiliary spaces  $\bar a,\bar b =
A_{1}\ldots A_{a}, B_{1}\ldots
B_{b}$  and the operator valued matrix with values in $\big(\CC^3\big) ^{\otimes^{a+b}}\otimes \Yan$,
\begin{equation}\label{product}
\ct_{\bar a, \bar b}(\bar u; \bar v)\ =\ \LL_{\bar a}(\bar u)\LL_{\bar b}(\bar v)
\,\ccR_{\bar b, \bar a}(\bar v; \bar u)
\ =\ \ccR_{\bar b, \bar a}(\bar v; \bar u)\,\LL_{\bar b}(\bar v)
\,\LL_{\bar a}(\bar u)\,,
\end{equation}
with
\ben\label{Rproduct}
 \ccR_{\bar b, \bar a}(\bar v; \bar u)=
{\prod^b_{i= 1}}\ {\prodl^a_{j = 1}}\,\,{R}_{B_iA_j}(v_i,u_j), \quad
\LL_{\bar a}(\bar u)=\prod_{i=1}^a\LL_{A_i}(u_i), \quad
\LL_{\bar b}(\bar v)=\prod_{i=1}^b\LL_{B_i}(v_i)\,.
\een
The last equality in \eqref{product} is a direct consequence of the RTT relation \eqref{RTT}.

The trace formula for a BV is given by
\begin{equation}\label{trace1}
\bbb^{a,b}(\bar\la; \bar\muu)
= \lambda^{-1}_2(\bar u)\,\lambda^{-1}_2(\bar v) {\rm tr}_{\bar a, \bar b}\ \big( \ct_{\bar a, \bar b}(\bar u; \bar v)
\,e_{21}^{\ot a}\ot e_{32}^{\ot b}\big)\rvec\,.
\end{equation}
This BV is symmetric in the set $\bar u$ and symmetric in the set $\bar v$ \cite{TarVar98}.

The normalization of $\bbb^{a,b}(\bar\la;\bar\mu)$ does not correspond to the one used in \cite{TarVar93,TarVar98}. To recover the original normalization one has to consider
\beq{eq:norm}
f(\bar v,\bar u)\,\lambda_2(\bar u)\,\lambda_2(\bar v)\bbb^{a,b}(\bar\la;\bar\mu).
\eeq

\section{Proofs \label{sec:proofs}}

This section collects the proofs of the statements formulated above. Let us comment
on the general strategy of the proofs and the order of presentation.

We first prove the equivalence of the explicit representations given in section~\ref{sect:expl} (section~\ref{prf:BV}).
Then we derive the multiple actions of $T_{13}(w)$, $T_{12}(w)$ and $T_{23}(w)$ on the vectors \eqref{BV-explicit-1}--\eqref{Or-form2}. The derivation is based only on the explicit formulas for these
vectors (sections~\ref{proofT13} and \ref{proofT12}). The proof of multiple actions of other generators requires the
iteration formula, therefore we postpone it to the end of the section and proceed to the proof of the recursions formulated
in section~\ref{sect:iter} (section~\ref{sect:iter-explicit}). Then, in section~\ref{prf:iter}, we show that the
trace formula \eqref{trace1} for BV obtained in \cite{TarVar98} implies the same recursions. This proves that the
explicit representations \eqref{BV-explicit-1}--\eqref{Or-form2} do define BV. In section~\ref{prf:phi} we show
explicitly the action of the morphism $\vph$ on BV. Finally, in section~\ref{proofT22} we complete the proofs
of multiple actions given in section~\ref{sec:TijBV}. The proofs are all similar and are based on the iteration formula. Therefore,
as an example, we restrict ourselves to a detailed consideration
of the multiple action of the operator $T_{22}(w)$.

\subsection{Equivalence of the different explicit expressions\label{prf:BV}}

We  prove that all the explicit formulas are equivalent.

As we have mentioned already the action of the morphism $\vph$ relates \eqref{BV-explicit-1} to \eqref{BV-explicit-2}, and  \eqref{Or-form} to  \eqref{Or-form2}. Hence, it remains to prove the equivalence between \eqref{Or-form} and \eqref{BV-explicit-1}.
In order to do this we consider
 \beq{G-def}
G=\sum \Izer_{n_1}(\bmu_{\so}|\bla_{\so})f(\bla_{\so},\bla_{\st})
T_{13}(\bla_{\so})T_{12}(\bla_{\st}),
\eeq
and substitute here \eqref{two}
with $i=1$, $j=3$, $k=2$, $\bar y=\bla_{\so}$, $\bar x=\bla_{\st}$ and $\bar w=\bla$. Then
 \beq{G1}
G=\sum \Izer_{n_1}(\bmu_{\so}|\bla_{\so})f(\bla_{\so},\bla_{\st})(-1)^{n_2}
\Izer_{n_2}(\bla_{\st}|\bar u_{\sttt}+c)T_{12}(\bar u_{\sttt})T_{13}(\bar u_{\stt})f(\bar u_{\sttt},\bar u_{\stt}).
\eeq
The two types of partitions of the set $\bla$, namely $\bla \Rightarrow \{\bla_{\so},\bla_{\st}\}$ and $\bla \Rightarrow \{\bar u_{\stt},\bar u_{\sttt}\}$, are
independent except that $\#\bla_{\so}=\#\bar u_{\stt}=n_1$ and $\#\bla_{\st}=\#\bar u_{\sttt}=n_2$. Hence, we can sum up over the partitions $\bla \Rightarrow\{\bla_{\so},\bla_{\st}\}$ via Lemma \ref{main-ident}. We obtain
 \beq{G2}
G=\sum (-1)^{n_1+n_2} f(\bmu_{\so},\bla)\Izer_{n_1+n_2}(\bla|\bmu_{\so}+c,\bar u_{\sttt}+c)T_{12}(\bar u_{\sttt})T_{13}(\bar u_{\stt})f(\bar u_{\sttt},\bar u_{\stt}).
\eeq
Now we apply \eqref{Red-K}, \eqref{K-K}. This gives
 \beq{G3}
G=\sum f(\bmu_{\so},\bar u_{\sttt})\Izer_{n_1}(\bmu_{\so}|\bar u_{\stt})T_{12}(\bar u_{\sttt})T_{13}(\bar u_{\stt})f(\bar u_{\sttt},\bar u_{\stt}).
\eeq
It remains to re-name $\bar u_{\stt}=\bla_{\so}$, $\bar u_{\sttt}=\bla_{\st}$ and to substitute this result into \eqref{Or-form} to recover \eqref{BV-explicit-1}.

\subsection{Action of $\LL_{13}(\bar w)$\label{proofT13}}

Consider a BV $\mathbb{B}^{a+1,b+1}(\bar\eta;\bar\xi)$ with $\{\bla, w\}=\bar\eta$ and
$\{\bmu, w\}=\bar\xi$. Due to \eqref{Or-form} we have
\ben
\label{Or-form1}
\bbb^{a+1,b+1}(\bar\eta;\bar\xi) =\sum \frac{\Izer_{k}(\bar\xi_{\so}|\bar\eta_{\so})}{\as_2(\bar \xi_{\st})\as_2(\bar \eta)}
\frac{f(\bar\xi_{\st},\bar\xi_{\so})f(\bar\eta_{\so},\bar\eta_{\st})}{f(\bar\xi,\bar\eta)}\,
T_{13}(\bar\eta_{\so})T_{12}(\bar\eta_{\st})T_{23}(\bar\xi_{\st})|0\rangle.
\een
The product $f^{-1}(\bar\xi,\bar\eta)$ contains vanishing factor $f^{-1}(w,w)$, which can be compensated only
by the pole of $\Izer_{k}(\bar\xi_{\so}|\bar\eta_{\so})$. Therefore we obtain a non-vanishing contribution to the BV
only if $w\in\bar\xi_{\so}$ and $w\in\bar\eta_{\so}$. Then we can set
\ben\label{Partitions}
\begin{array}{ll}
\{\bla_{\so}, w\}=\bar\eta_{\so},& \bla_{\st}=\bar\eta_{\st},\\
\{\bmu_{\so}, w\}=\bar\xi_{\so},& \bmu_{\st}=\bar\xi_{\st}.
\end{array}
\een
Substituting \eqref{Partitions} into \eqref{Or-form1} and using \eqref{Resid-K} for $\Izer_{k}(\bar\xi_{\so}|\bar\eta_{\so})$
we immediately arrive at \eqref{act13} with $n=1$. Trivial recursion over $n=\#\bar w$ ends the proof.

\subsection{\label{proofT12} Action of $T_{12}(\bar w)$ and $T_{23}(\bar w)$}

We first prove the formula for $T_{12}(\bar w)$, starting from the expression \eqref{BV-explicit-1}.

Let $\bar w$, $\bar\eta$, and $\bar\xi$ be arbitrary complex numbers with
$\#\bar w=n$, $\#\bar\eta=a+n$, and $\#\bar\xi=b+n$. Consider the following combination
of BV:
 \beq{act12G}
 G(\bar w)=(-1)^n\sum
 f(\bar\xi_{\st},\bar\xi_{\so})\Izer_n(\bar\xi_{\so}|\bar w+c)
 \mathbb{B}^{a+n,b}(\bar\eta;\bar\xi_{\st}).
 \eeq
The sum is taken over partitions $\bar\xi\Rightarrow\{\bar\xi_{\so},\bar\xi_{\st}\}$ with $\#\bar\xi_{\so}=n$. Substituting here \eqref{BV-explicit-1} and using also \eqref{Red-K}
we obtain
 \begin{multline}\label{1}
 G(\bar w)=\sum(-1)^{n+k}
 f(\bar\xi_{\st},\bar\xi_{\so})\Izer_n(\bar\xi_{\so}|\bar w+c)
\Izer_{k}(\bar\eta_{\so}-c|\bar\xi_{\rm i})f^{-1}(\bar\xi_{\rm ii},\bar\eta)f(\bar\xi_{\rm ii},\bar\xi_{\rm i})
f(\bar\eta_{\st},\bar\eta_{\so})\\
\times\lambda_2^{-1}(\bar\eta)\,\lambda_2^{-1}(\bar\xi_{\rm ii})
T_{12}(\bar\eta_{\st})T_{13}(\bar\eta_{\so})T_{23}(\bar\xi_{\rm ii})|0\rangle.
\end{multline}
Here the subset $\bar\xi_{\st}$ is divided into sub-subsets: $\bar\xi_{\st}\Rightarrow\{\bar\xi_{\rm i},\bar\xi_{\rm ii}\}$.
The sum is taken with respect to all partitions described above.

Let
$\{\bar\xi_{\so},\bar\xi_{\rm i}\}=\bar\xi_0$. Then
 \beq{prosto}
 f(\bar\xi_{\st},\bar\xi_{\so})f(\bar\xi_{\rm ii},\bar\xi_{\rm i})=
 f(\bar\xi_{\rm i},\bar\xi_{\so})f(\bar\xi_{\rm ii},\bar\xi_0),
 \eeq
and substituting this into \eqref{1} we obtain
 \begin{multline}\label{2}
 G(\bar w)=\sum(-1)^{n+k}
 f(\bar\xi_{\rm i},\bar\xi_{\so})\Izer_n(\bar\xi_{\so}|\bar w+c)
\Izer_{k}(\bar\eta_{\so}-c|\bar\xi_{\rm i})f^{-1}(\bar\xi_{\rm ii},\bar\eta)f(\bar\xi_{\rm ii},\bar\xi_0)
f(\bar\eta_{\st},\bar\eta_{\so})\\
\times\lambda_2^{-1}(\bar\eta)\,\lambda_2^{-1}(\bar\xi_{\rm ii})
T_{12}(\bar\eta_{\st})T_{13}(\bar\eta_{\so})T_{23}(\bar\xi_{\rm ii})|0\rangle.
\end{multline}
The sum over partitions $\bar\xi_0\Rightarrow\{\bar\xi_{\rm i},\bar\xi_{\so}\}$ can be computed
via \eqref{Red-K}, \eqref{Sym-Part-old2}
\beq{sum}
\sum_{\bar\xi_0=\{\bar\xi_{\rm i},\bar\xi_{\so}\}} f(\bar\xi_{\rm i},\bar\xi_{\so})\Izer_n(\bar\xi_{\so}|\bar w+c)
\Izer_{k}(\bar\eta_{\so}-c|\bar\xi_{\rm i})=(-1)^kf^{-1}(\bar\xi_0,\bar\eta_{\so})
\Izer_{n+k}(\bar\xi_0|\bar\eta_{\so},\bar w+c).
\eeq
Thus, we have
\begin{multline}\label{3}
 G(\bar w)=\sum(-1)^{n}
 \Izer_{n+k}(\bar\xi_0|\bar\eta_{\so},\bar w+c)f^{-1}(\bar\xi_0,\bar\eta_{\so})f^{-1}(\bar\xi_{\rm ii},\bar\eta)f(\bar\xi_{\rm ii},\bar\xi_0)
f(\bar\eta_{\st},\bar\eta_{\so})\\
\times\lambda_2^{-1}(\bar\eta)\,\lambda_2^{-1}(\bar\xi_{\rm ii})
T_{12}(\bar\eta_{\st})T_{13}(\bar\eta_{\so})T_{23}(\bar\xi_{\rm ii})|0\rangle.
\end{multline}
The sum is taken over partitions $\bar\eta\Rightarrow\{\bar\eta_{\so},\bar\eta_{\st}\}$ (as it was from the very
beginning) and  $\bar\xi\Rightarrow\{\bar\xi_{\rm ii},\bar\xi_0\}$.

Up to now $\bar w$, $\bar\eta$, and $\bar\xi$ were arbitrary complex. Let now
$\{\bar u,\bar w\}=\bar\eta$ and $\{\bar v,\bar w\}=\bar\xi$. Then $\bar w\subset\bar\xi_0$,
otherwise due to the factor $f^{-1}(\bar\xi_{\rm ii},\bar\eta)$ we obtain a vanishing contribution. Then
$\bar w\subset\bar\eta_{\st}$, otherwise due to the factor $f^{-1}(\bar\xi_{0},\bar\eta_{\so})$ we
again obtain zero. Thus, we can set
 \beq{raxb}
 \begin{array}{l}
\{\bar w, \bar v_{\so}\}= \bar\xi_0,\qquad \bar\xi_{\rm ii}=\bar v_{\st},\\
\{\bar w, \bar u_{\st}\}= \bar\eta_{\st},\qquad \bar\eta_{\so}=\bar u_{\so}.
 \end{array}
 \eeq
Substituting this into \eqref{3} and using \eqref{Red-K}, \eqref{K-K} we arrive at
 \begin{multline}\label{Alt-form-2}
G(\bar w)=
\sum \Izer_{k}(\bar v_{\so} |\bar u_{\so})f^{-1}(\bar v_{\st},\bar u)f^{-1}(\bar v_{\so},\bar u_{\so})
f(\bar v_{\st},\bar v_{\so})f(\bar u_{\st},\bar u_{\so})\\
\times\lambda_2^{-1}(\bar\eta)\,\lambda_2^{-1}(\bar v_{\st})
T_{12}(\bar w)T_{12}(\bar u_{\st})T_{13}(\bar u_{\so})T_{23}(\bar v_{\st})|0\rangle
=
\lambda_2^{-1}(\bar w)\,T_{12}(\bar w)\mathbb{B}^{a,b}(\bar u;\bar v).
\end{multline}
Thus, we have proved that
 \beq{act12-doc}
 T_{12}(\bar w)\mathbb{B}^{a,b}(\bar u;\bar v)=(-1)^n\lambda_2(\bar w)\,\sum
 f(\bar\xi_{\st},\bar\xi_{\so})\Izer_n(\bar\xi_{\so}|\bar w+c)
 \mathbb{B}^{a+n,b}(\bar\eta;\bar\xi_{\st}),
 \eeq
where $\{\bar u,\bar w\}=\bar\eta$ and $\{\bar v,\bar w\}=\bar\xi$.

Applying the morphism $\vph$ to the relation \eqref{act12}, we obtain \eqref{act23}.

\subsection{\label{sect:iter-explicit}Proof of the iteration formulas from explicit representations}

The iteration formulas \eqref{iter1}, \eqref{iter2} immediately follow from the actions \eqref{act12}, \eqref{act23} derived
above.
For instance, let us replace $\mathbb{B}^{a,b}(\bla;\bmu)$ in \eqref{act12} by $\mathbb{B}^{a-1,b}(\bar u_k;\bmu)$,
where $u_k$ is an arbitrary element of the set $\bar u$. Applying the operator $\LL_{12}(u_k)$ to this vector via
\eqref{act12} we find
 \beq{act12-1}
 T_{12}(u_k)\mathbb{B}^{a-1,b}(\bar u_k;\bmu)=\lambda_2(u_k)\,\sum
 f(\bar\xi_{\st},\bar\xi_{\so})h^{-1}(u_k,\bar\xi_{\so})\,
 \mathbb{B}^{a,b}(\bar u;\bar\xi_{\st}),
 \eeq
where $\{\bar v, u_k\}=\bar\xi\Rightarrow\{\bar\xi_{\so},\bar\xi_{\st}\}$ and the subset $\bar\xi_{\so}$ consists of one element: $\#\bar\xi_{\so}=1$. Setting here $\bar\xi_{\so}=u_k$ we reproduce the l.h.s. of \eqref{iter1}. Setting $\bar\xi_{\so}=v_i$,
$i=1,\dots,b$, and using $\lambda_2(u_k)\mathbb{B}^{a,b}(\bar u;\{\bar v_i,u_k\})=\LL_{13}(u_k)
\mathbb{B}^{a-1,b-1}(\bar u_k;\bar v_i)$ we obtain the sum of terms in the r.h.s. of \eqref{iter1}.

Similarly equation \eqref{act23} produces the iteration formula \eqref{iter2}.

\subsection{Equivalence of the iteration formulas with the trace formula \label{prf:iter}}

Let us consider the trace formula (\ref{trace1}). We single out the trace over the first auxiliary space to obtain the iteration formula
\begin{equation}\label{First-sp}
\bbb^{a,b}(\bar\la;\bar\muu)
=\lambda_2^{-1}(\bar u)\lambda_2^{-1}(\bar v){\rm tr}_{A_1}\Big(T_{A_1}(u_1) {\rm tr}_{\bar a_1,\bar b} \big(\ct_{\bar a_1,\bar b}(\bar\la_1;\bar\muu)\,\prod_{l=1}^bR_{B_lA_1}(v_l,u_1)
\, e_{21}^{\ot a}\ot e_{32}^{\ot b}\big)\Big) \, \rvec,
\end{equation}
and factor out the first auxiliary space from the product of $R$-matrices
$$
\prod_{i=1}^bR_{B_i,A_1}(v_i,u_1)e_{21}^{\ot a}\ot e_{32}^{\ot b}
=
f^{-1}(\bar v,u_1)e_{21}^{\ot a}\ot e_{32}^{\ot b}+\sum_{j=1}^{b} g(v_j,u_1)\prod_{l=j}^{b} f^{-1}(v_l,u_1)e_{31}\ot E_j,\nonumber
$$
with $E_j=e_{21}^{\ot a-1}\ot e_{32}^{\ot j-1}\otimes e_{22} \otimes e_{32}^{\ot b-j}$. Taking the trace over the first space we obtain
\ben\label{dev1}
\bbb^{a,b}(\bar\la;\bar\muu)
&=&  \lambda_2^{-1}(u_1)f^{-1}(\bar v,u_1)T_{12}(u_1)\bbb^{a-1,b}(\bar\la_1;\bar\muu)\\
&& + \lambda^{-1}_2(\bar u)\lambda^{-1}_2(\bar v)\sum_{j=1}^{b}  g(v_j,u_1) \prod_{l=j}^{b} f^{-1}(v_l,u_1)T_{13}(u_1) X_j,\nonumber
\een
with $X_j=
{\rm tr}_{\bar a_1,\bar b} \big(\ct_{\bar a_1,\bar b}(\bar\la_1;\bar\muu)\,E_j\big)\rvec$.
To compute the sum in \eqref{dev1}, we show in appendix \ref{app:X} that $X_j$ obeys the relation
\ben
X_j+\sum_{k=j+1}^bg(v_k,v_j) \prod_{l=j+1}^{k-1}f(v_l,v_j)\,X_k=\lambda_2(\bar u_1)\lambda_2(\bar v)\prod_{l=j+1}^{b}f(v_l,v_j)\bbb^{a-1,b-1}(\bar\la_1;\bar\muu_j).
\label{eq:triangX}
\een
This equation shows that any $X_j$ can be written as a linear combination of  $ \bbb^{a-1,b-1}(\bar\la_1;\bar\muu_k)$ with $k\geq j$.

If we look for $\bbb^{a-1,b-1}(\bar\la_1;\bar\muu_1)$, we see that it can appear only in $X_1$. We have
\ben
X_1=\lambda_2(\bar v)\lambda_2(\bar u_1)f(\bar v_1,v_1) \bbb^{a-1,b-1}(\bar\la_1;\bar\muu_1)+... ,
\een
where dots stand for terms containing $\bbb^{a-1,b-1}(\bar\la_1;\bar\muu_k)$, $k>1$.
It follows that  (\ref{dev1}) can be rewritten as
\ben
\bbb^{a,b}(\bar\la;\bar\muu)
&=&  \lambda_2^{-1}(u_1)f^{-1}(\bar v,u_1)T_{12}(u_1)\bbb^{a-1,b}(\bar\la_1;\bar\muu)
\nonu
&&+\lambda^{-1}_2(u_1)f^{-1}(\bar v,u_1)f(\bar v_1,v_1)g(v_1,u_1)T_{13}(u_1)\bbb^{a-1,b-1}(\bar\la_1;\bar\muu_1)+... ,
\een
where dots still stand for terms containing $\bbb^{a-1,b-1}(\bar\la_1;\bar\muu_k)$, $k>1$.

Since
$ \bbb^{a,b}(\bar\la;\bar\muu)$ is symmetric in $\bar v$, all the other terms  must share the same form, and we get \eqref{iter1}.

\medskip

Applying the morphism $\vph$ to \eqref{iter1}, one gets \eqref{iter2}.

\subsection{Action  of the morphism $\vph$ on Bethe vectors\label{prf:phi}}

We show the explicit action of the morphism $\vph$ via the iteration formulas. It is clear from the initial condition \eqref{BV-initial} that $\mathbb{B}^{a,0}(\bar u)$ and $\mathbb{B}^{0,b}(\bar v)$ are related by $\vph$. Then application of $\vph$ to the iteration formula
\eqref{iter1} leads to (using the induction hypothesis):
 \begin{multline}\label{App-vph}
\lambda_2(-u_k)f(\bar v, u_k)\, \vph\big( \mathbb{B}^{a,b}(\bar u;\bar v)\big)=
T_{23}(-u_k) \mathbb{B}^{b,a-1}(-\bar v;-\bar u_k)
\\
+\sum_{i=1}^b g( v_i,u_k)f(\bar v_i, v_i)  T_{13}(- u_k)\mathbb{B}^{b-1,a-1}(-\bar v_i; -\bar u_k)\,.
%\nonumber
\end{multline}
Setting $a'=b$, $b'=a$, $\bar u'=-\bar v$ and $\bar v'=-\bar u$, one obtains
 \begin{multline}
 \label{phi-iter2}
\lambda_2(u'_k)f(\bar v', u'_k)\,\vph\big( \mathbb{B}^{b',a'}(-\bar v';-\bar u')\big)=
T_{23}(v'_k) \mathbb{B}^{a',b'-1}(\bar u';\bar v_k) \\
+ \sum_{i=1}^{a'} g( v'_k,u'_i)f(u'_i,-\bar u'_i) T_{13}(v'_k)\mathbb{B}^{a'-1,b'-1}(\bar u'_i; \bar v'_k),
%\nonumber
\end{multline}
where we have used $f(-\bar v,-\bar u)=f(\bar u,\bar v)$ and $g(-u_i,-v_k)=g(v_k,u_i)$.

One recognizes in the r.h.s. of \eqref{phi-iter2} the iteration formula \eqref{iter2} for $\mathbb{B}^{a',b'}(\bar u';\bar v')$.

%%%%%%%%%%%%BEGIN

\subsection{Action of $\LL_{ij}$, $i\geq j$ \label{proofT22}}

The actions of other operators $\LL_{ij}$ with $i\ge j$ on BV also can be proved by the use of 
explicit formulas for the BV. However the corresponding proofs are quite cumbersome. Instead one can use
the recursion \eqref{iter1} (resp. \eqref{iter2}) and prove equations \eqref{act22}--\eqref{act31}
via induction over $a$ (resp. $b$). As an example we give the detailed proof of the action \eqref{act22}.

We first check that the equation \eqref{act22} is valid for $a=0$ and arbitrary $b$.
Setting $\bar u=\emptyset$ in \eqref{act22} we obtain $\bar\eta_{\so}=\bar w$ and
$\bar\eta_{\st}=\emptyset$. Hence, using $\Izer_n(\bar w|\bar w+c)=(-1)^n$, we obtain
 \begin{equation}\label{act22-0}
 \LL_{22}(\bar w)\mathbb{B}^{0,b}(\bar v)=\lambda_2(\bar w)(-1)^n\,\sum
f(\bar\xi_{\st},\bar\xi_{\so})
  \Izer_n(\bar\xi_{\so}|\bar w+c)\,\mathbb{B}^{0,b}(\bar\xi_{\st}).
 \end{equation}
On the other hand, due to \eqref{BV-initial}
\begin{equation}\label{B0b}
\mathbb{B}^{0,b}(\bar v)=\lambda_2^{-1}(\bar v)\LL_{23}(\bar v)|0\rangle.
\end{equation}
Using \eqref{two} we reproduce \eqref{act22-0}.

Assuming now that \eqref{act22} holds for $\mathbb{B}^{a,b}(\bar u;\bar v)$ with
fixed $a$ and $\#\bar w=n=1$, we prove that the same action is valid for  $\mathbb{B}^{a+1,b}(\{\bar u,x\};\bar v)$ using
\eqref{iter1}. It is clear that for this we have to calculate the successive action of the operators
$\LL_{22}(w)\LL_{12}(x)$ and $\LL_{22}(w)\LL_{13}(x)$ on the BV.

We first derive the action  $\LL_{22}(w)\LL_{13}(x)$ on the BV of the form $\mathbb{B}^{a,b-1}(\bar u;\bar v)$.
Using \eqref{Ycom2} we obtain
\begin{multline}\label{T22-13}
\LL_{22}(w)\LL_{13}(x)\mathbb{B}^{a,b-1}(\bar u;\bar v) =\LL_{13}(x)\LL_{22}(w)\mathbb{B}^{a,b-1}(\bar u;\bar v)\\
 + g(y,x)\Big(\LL_{23}(w)\,\LL_{12}(x)-\LL_{23}(x)\,\LL_{12}(w)\Big)\mathbb{B}^{a,b-1}(\bar u;\bar v).
\end{multline}
The action of $\LL_{22}(w)$ on $\mathbb{B}^{a,b-1}(\bar u;\bar v)$ is known due to the induction
assumption, the actions of $\LL_{13}$, $\LL_{12}$, and $\LL_{23}$ are known for arbitrary BV.

Let us compute the action $\LL_{23}(x)\LL_{12}(w)$. Using successively \eqref{act12}, \eqref{act23} we find
 \begin{multline}\label{SA2312-3}
 \LL_{23}(x)\LL_{12}(w)\mathbb{B}^{a,b-1}(\bar u;\bar v)=\lambda_2(x)\lambda_2(w)\,\sum
\frac{ f(\bar\xi_{\st},\bar\xi_{\so})f(\bar\eta_{\so},\bar\eta_{\st})}{f(x,\bar\xi_{\so})}\\
 \times\Izer_1(\bar\xi_{\so}|w+c)\Izer_1(x|\bar\eta_{\so}+c)\,
 \mathbb{B}^{a+1,b}(\bar\eta_{\st};\bar\xi_{\st}).
 \end{multline}
Here the sum is taken over partitions of the sets: $\{\bar u,w,x\}=\bar\eta\Rightarrow\{\bar\eta_{\so},\bar\eta_{\st}\}$ with $\#\bar\eta_{\so}=1$; $\{\bar v,w,x\}=\bar\xi\Rightarrow\{\bar\xi_{\so},\bar\xi_{\st}\}$ with $\#\bar\xi_{\so}=1$.

The action of $\LL_{13}(x)\LL_{22}(w)$ reads
 \begin{multline}\label{SA1322-2}
 \LL_{13}(x)\LL_{22}(w)\mathbb{B}^{a,b-1}(\bla;\bmu)=\lambda_2(x)\lambda_2(w)\,\sum
\frac{f(\bar\xi_{\st},\bar\xi_{\so}) f(\bar\eta_{\so},\bar\eta_{\st})}
{f(x,\bar\xi_{\so}) f(\bar\eta_{\so},x)}\\
\times   \Izer_1(\bar\xi_{\so}|y+c)
 \Izer_1(w|\bar\eta_{\so}+c)\,\mathbb{B}^{a+1,b}(\bar\eta_{\st};\bar\xi_{\st}).
 \end{multline}
The notations are the same as in \eqref{SA2312-3}.
It remains to substitute \eqref{SA2312-3} and \eqref{SA1322-2} into \eqref{T22-13}. It is straightforward to check
that we obtain
 \begin{multline}\label{SA2213}
 \LL_{22}(w)\LL_{13}(x)\mathbb{B}^{a,b-1}(\bla;\bmu)=\lambda_2(x)\lambda_2(w)\,\sum
f(\bar\xi_{\st},\bar\xi_{\so}) f(\bar\eta_{\so},\bar\eta_{\st})
\\
\times   \Izer_1(\bar\xi_{\so}|w+c)
 \Izer_1(w|\bar\eta_{\so}+c)\,\mathbb{B}^{a+1,b}(\bar\eta_{\st};\bar\xi_{\st}).
 \end{multline}

Similarly, using
 \beq{T22T12-n1}
\LL_{22}(w)\LL_{12}(x)=f(w,x)\LL_{12}(x)\LL_{22}(w)+g(w,x)\LL_{12}(w)\LL_{22}(x),
\eeq
one can find that
 \begin{multline}\label{SA-act22-7}
\LL_{22}(w)\LL_{12}(x)\mathbb{B}^{a,b}(\bla;\bmu)=-\lambda_2(x)\lambda_2(w)\sum
 f(\bar\xi_{\st},\bar\xi_{\so}) f(\bar\eta_{\so},\bar\eta_{\st})\\
 \times
\Izer_{1}(w|\bar\eta_{\so}+c)\Izer_{2}(\bar\xi_{\so}|\{x+c, w+c\})
 \,\mathbb{B}^{a+1,b}(\bar\eta_{\st};\bar\xi_{\st}),
 \end{multline}
where the sum is taken over partitions:
$\{\bar u,w,x\}=\bar\eta\Rightarrow\{\bar\eta_{\so},
\bar\eta_{\st}$\} with $\#\bar\eta_{\so}=1$; $\{\bar v,w,x\}=\bar\xi\Rightarrow\{\bar\xi_{\so},\bar\xi_{\st}\}$
with $\#\bar\xi_{\so}=2$.

Now everything is ready for the calculation of the action of $\LL_{22}(w)$ on BV of the form $\mathbb{B}^{a+1,b}(\{\bar u,x\};\bar v)$. It follows from \eqref{act12} at $n=1$ that
 \begin{equation}\label{Act22-1}
 \lambda_2(x)f(\bar v, x)\LL_{22}(w)\mathbb{B}^{a+1,b}(\{\bar u,x\};\bar v)=
 \LL_{22}(w)\LL_{12}(x)\mathbb{B}^{a,b}(\bar u;\bar v)+G(x,w),
 \end{equation}
where
\begin{equation}\label{Gxy}
G(x,w)=\lambda_2(x){\sum}'f(\bar\xi_{\st},\bar\xi_{\so})\Izer_1(\bar\xi_{\so}|x+c)
\LL_{22}(w)\mathbb{B}^{a+1,b}(\{\bar u,x\};\bar\xi_{\st}).
\end{equation}
Here the symbol $\sum'$ means that the sum is taken over partitions of the set $\{\bar v, x\}=\bar\xi\Rightarrow
\{\bar\xi_{\so},\bar\xi_{\st}\}$ with $\#\bar\xi_{\so}=1$ and the restriction $\bar\xi_{\so}\ne x$. Observe that in this case
the subset $\bar\xi_{\st}$ is $\{\bar v_i,x\}=\bar\xi_{\st}$, where $i=1,\dots,b$. Thus, the BV in \eqref{Gxy} has
the form $\mathbb{B}^{a+1,b}(\{\bar u,x\};\{\bar v_i,x\})$, and hence, it can be presented as the result of the action
$\LL_{13}(x)$: $\mathbb{B}^{a+1,b}(\{\bar u,x\};\{\bar v_i,x\})=\lambda_2^{-1}(x)\LL_{13}(x)\mathbb{B}^{a,b-1}(\bar u; \bar v_i)$.
Therefore  we can use \eqref{SA2213} for the evaluation of the action of $\LL_{22}$ on such vector. We obtain
\begin{multline}\label{Gxy1}
G(x,w)=\lambda_2(x)\lambda_2(w){\sum}'f(\bar\xi_{\st},\bar\xi_{\so})f(\bar\eta_{\so},\bar\eta_{\st})
f(\bar\xi_{\rm ii},\bar\xi_{\rm i})\\
\times\Izer_1(\bar\xi_{\so}|x+c)\Izer_1(\bar\xi_{\rm i}|w+c)
\Izer_1(w|\bar\eta_{\so})
\mathbb{B}^{a+1,b}(\bar\eta_{\st};\bar\xi_{\rm ii}).
\end{multline}
Here there is the sum over additional partitions, the set $\{\bar u,x,w\}=\bar\eta\Rightarrow\{\bar\eta_{\so},\bar\eta_{\st}\}$; the set $\{\bar\xi_{\st},w\}\Rightarrow\{\bar\xi_{\rm i},\bar\xi_{\rm ii}\}$. Thus, the set $\{\bar v,x,w\}$ is actually  divided into three subsets: $\{\bar v,x,w\}=\bar\xi\Rightarrow\{\bar\xi_{\so},\bar\xi_{\rm i},
\bar\xi_{\rm ii}\}$ with the conditions $\bar\xi_{\so}\ne w$ and $\bar\xi_{\so}\ne x$. Hereby
$\{\bar\xi_{\rm i},\bar\xi_{\rm ii}\}\setminus w=\bar\xi_{\st}$. Then
\begin{multline}\label{Gxy2}
G(x,w)=\lambda_2(x)\lambda_2(w){\sum}'\frac{f(\bar\xi_{\rm ii},\bar\xi_{\so})
f(\bar\xi_{\rm i},\bar\xi_{\so})f(\bar\xi_{\rm ii},\bar\xi_{\rm i})f(\bar\eta_{\so},\bar\eta_{\st})}
{f(w,\bar\xi_{\so})}\\
\times\Izer_1(\bar\xi_{\so}|x+c)\Izer_1(\bar\xi_{\rm i}|w+c)
\Izer_1(w|\bar\eta_{\so})
\mathbb{B}^{a+1,b}(\bar\eta_{\st};\bar\xi_{\rm ii}).
\end{multline}
We see that now the condition $\bar\xi_{\so}\ne w$
holds automatically due to the factor $f^{-1}(w,\bar\xi_{\so})$. In order to get rid of the restriction $\bar\xi_{\so}\ne x$
we present $G(x,w)$ as
\begin{equation}\label{presG}
G(x,w)=G_1(x,w)-G_2(x,w),
\end{equation}
where $G_1(x,w)$ is the sum \eqref{Gxy2} without any restriction, and $G_2(x,w)$ is the sum \eqref{Gxy2}
at $\bar\xi_{\so}=x$. Then we have for $G_1(x,w)$
\begin{multline}\label{Gxy3}
G(x,w)=-\lambda_2(x)\lambda_2(w){\sum}\frac{f(\bar\xi_{\rm ii},\bar\xi_{0})
f(\bar\eta_{\so},\bar\eta_{\st})}
{f(w,\bar\xi_{0})}\Izer_1(w|\bar\eta_{\so})\\
\times\Izer_1(\bar\xi_{\so}|x+c)\Izer_1(w|\bar\xi_{\rm i})
f(\bar\xi_{\rm i},\bar\xi_{\so})
\mathbb{B}^{a+1,b}(\bar\eta_{\st};\bar\xi_{\rm ii}),
\end{multline}
where we introduced
$\{\bar\xi_{\so},\bar\xi_{\rm i}\}=\bar\xi_{0}$ and used \eqref{Red-K} for $\Izer_1(\bar\xi_{\rm i}|w+c)$.
Now we can apply Lemma~\ref{main-ident} to the second line of \eqref{Gxy3}:
\begin{equation}\label{S-l}
\sum\Izer_1(\bar\xi_{\so}|x+c)\Izer_1(w|\bar\xi_{\rm i})
f(\bar\xi_{\rm i},\bar\xi_{\so})=-f(w,\bar\xi_{0})\Izer_2(\bar\xi_{0}|\{x+c,w+c\}).
\end{equation}
Substituting this into \eqref{Gxy3} and comparing with \eqref{SA-act22-7}  we conclude that
\begin{equation}\label{res-G1}
G_1(x,w)=-\LL_{22}(w)\LL_{12}(x)\mathbb{B}^{a,b}(\bar u;\bar v).
\end{equation}

On the other hand setting in \eqref{Gxy2} $\bar\xi_{\so}=x$  we
obtain
\begin{multline}\label{Gxy4}
G_2(x,w)=-\lambda_2(x)\lambda_2(w){\sum}\frac{f(\bar\xi_{\rm ii},x)
f(\bar\xi_{\rm i},x)f(\bar\xi_{\rm ii},\bar\xi_{\rm i})f(\bar\eta_{\so},\bar\eta_{\st})}
{f(w,x)}\\
\times\Izer_1(\bar\xi_{\rm i}|w+c)
\Izer_1(w|\bar\eta_{\so})
\mathbb{B}^{a+1,b}(\bar\eta_{\st};\bar\xi_{\rm ii}),
\end{multline}
where the sum is taken over partitions of the sets: $\{\bar v,w\}\!=\!\bar\xi\Rightarrow\{\bar\xi_{\rm i},\bar\xi_{\rm ii}\}$;
$\{\bar u,x,w\}\!=\!\bar\eta\Rightarrow\{\bar\eta_{\so},\bar\eta_{\st}\}$. It remains to observe that
$f(\bar\xi_{\rm ii},x)f(\bar\xi_{\rm i},x)=f(\bar v,x)f(w,x)$, and we find
\begin{equation}\label{Gxy5}
G_2(x,w)=-\lambda_2(x)\lambda_2(w)f(\bar v,x){\sum}f(\bar\xi_{\rm ii},\bar\xi_{\rm i})f(\bar\eta_{\so},\bar\eta_{\st})
\Izer_1(\bar\xi_{\rm i}|w+c)\Izer_1(w|\bar\eta_{\so})
\mathbb{B}^{a+1,b}(\bar\eta_{\st};\bar\xi_{\rm ii}).
\end{equation}
Finally  substituting \eqref{Gxy5}, \eqref{res-G1}, and \eqref{presG} into \eqref{Act22-1}  we obtain
\begin{equation}\label{Gxy6}
\LL_{22}(w)\mathbb{B}^{a+1,b}(\{\bar u,x\};\bar v)=\lambda_2(w){\sum}f(\bar\xi_{\rm ii},\bar\xi_{\rm i})f(\bar\eta_{\so},\bar\eta_{\st})
\Izer_1(\bar\xi_{\rm i}|w+c)\Izer_1(w|\bar\eta_{\so})
\mathbb{B}^{a+1,b}(\bar\eta_{\st};\bar\xi_{\rm ii}).
\end{equation}
Thus, the step of induction over $a$ is completed, and it remains to prove that the action
\eqref{act22} holds for $\#\bar w=n\ge 1$ as well. We again
use  induction. Suppose that \eqref{act22} is valid for the multiple action of
$\LL_{22}(\bar w_n)$. Then similarly to the calculations described above
we obtain for successive action of $\LL_{22}(\bar w_n)$ and $\LL_{22}(w_n)$
 \begin{multline}\label{act22-d2}
 \LL_{22}(\bar w)\mathbb{B}^{a,b}(\bla;\bmu)=\lambda_2(\bar w)\,\sum
\frac{f(\bar\xi_{\rm i},\bar\xi_{\so})f(\bar\xi_{\rm ii},\bar\xi_{0})
 f(\bar\eta_{\so},\bar\eta_{\rm i}) f(\bar\eta_{0},\bar\eta_{\rm ii})}
 {f(w_n,\bar\xi_{\so})f(\bar\eta_{\so},w_n)}   \\
 \times
   \Izer_{n-1}(\bar\xi_{\so}|\bar w_n+c)\Izer_1(\bar\xi_{\rm i}| w_n+c)
 \Izer_{n-1}(\bar w_n|\bar\eta_{\so}+c)  \Izer_1(w_n|\bar\eta_{\rm i}+c)\,\mathbb{B}^{a,b}(\bar\eta_{\rm ii};\bar\xi_{\rm ii}).
 \end{multline}
Here we take the sum over partitions, the set $\{\bar u,\bar w\}=\bar\eta\Rightarrow\{\bar\eta_{\so},\bar\eta_{\rm i},\bar\eta_{\rm ii}\}$
and $\{\bar v,\bar w\}=\bar\xi\Rightarrow\{\bar\xi_{\so},\bar\xi_{\rm i},\bar\xi_{\rm ii}\}$.
 Hereby $\#\bar\eta_{\rm i}=\#\bar\xi_{\rm i}=1$ and $\#\bar\eta_{\so}=\#\bar\xi_{\so}=n-1$.
We also have introduced
$\{\bar\eta_{\rm i},\bar\eta_{\so}\}=\bar\eta_{0}$ and  $\{\bar\xi_{\rm i},\bar\xi_{\so}\}=\bar\xi_{0}$.

Transforming the coefficients $\Izer_1$ via \eqref{Red-K} we obtain
 \begin{multline}\label{act22-d3}
 \LL_{22}(\bar w)\mathbb{B}^{a,b}(\bla;\bmu)=\lambda_2(\bar w)\,\sum
\frac{f(\bar\xi_{\rm i},\bar\xi_{\so})f(\bar\xi_{\rm ii},\bar\xi_{0})
 f(\bar\eta_{\so},\bar\eta_{\rm i}) f(\bar\eta_{0},\bar\eta_{\rm ii})}
 {f(w_n,\bar\xi_{0})f(\bar\eta_{0},w_n)}   \\
 \times
   \Izer_{n-1}(\bar\xi_{\so}|\bar w_n+c)\Izer_1(w_n|\bar\xi_{\rm i})
 \Izer_{n-1}(\bar w_n|\bar\eta_{\so}+c)  \Izer_1(\bar\eta_{\rm i}|w_n)\,\mathbb{B}^{a,b}(\bar\eta_{\rm ii};\bar\xi_{\rm ii}),
 \end{multline}
and we can use \eqref{Sym-Part-old2} for the summation over partitions $\bar\eta_{0}\Rightarrow\{\bar\eta_{\rm i},\bar\eta_{\so}\}$ and  $\bar\xi_{0}\Rightarrow\{\bar\xi_{\rm i},\bar\xi_{\so}\}$. Then we immediately arrive at \eqref{act22} with $\#\bar w=n$.

The proofs of \eqref{act11}--\eqref{act31} can be given in the same manner. They are, however, rather
cumbersome, therefore we do not give them in order to lighten the presentation. We just remark that the morphism $\vph$
allows one to obtain \eqref{act33} and \eqref{act32} from \eqref{act11} and \eqref{act21} respectively.

\section*{Conclusion}

In this paper we have computed different explicit expressions for BV, and derived the actions of the monodromy matrix entries on them. This is a first step towards the
 calculation of form factors and correlation functions.  Indeed, to compute correlation functions or  form factors of some local operator from the ABA, one has to reconstruct the local operators with the $\cY(\mathfrak{gl}_3)$ generators entering the monodromy matrix. The solution of this problem is given in \cite{KitMaiT99,MaiTer00}. Then, the action of the monodromy matrix on BV (as performed in the present paper) reduces the problem to calculation of scalar products of BV.

In the case of integrable models associated with the Yangian $\cY(\mathfrak{gl}_2)$ or the quantum algebra $\cU_q(\widehat {\mathfrak{gl}}_2)$, the scalar products of two BV were considered in \cite{Kor82,IzeKor84,Kor84,Ize87,KorBogI93}. There, a formula involving  sums of products of two determinants was found.
It was shown in \cite{Sla89,Sla07} that if one of the BV is on-shell, then the scalar product can be written in terms of a single determinant. This single determinant formulation of the scalar product is essential in the explicit calculation of correlation functions and form factors for the related physical models.

In the case of the models described by the Yangian $\cY(\mathfrak{gl}_3)$, the structure of the scalar product of two BV was given in \cite{Res86}.  This general formula is very cumbersome. Taking into account that the
formulas of multiple actions of the monodromy matrix entries on BV are also quite complex,
one can expect that the resulting expressions for form factors will be unacceptable for their analysis.
However, there are several arguments that give hope of handling the problem.

First of all we would like to draw the reader's attention to the fact that the most cumbersome formulas  for
the action of  the low-triangular part of the monodromy matrix on the BV actually are not
necessary for the calculation of form factors. Indeed, the form factors of $T_{ij}(w)$ with $i>j$
are related with the form factors of $T_{ij}(w)$ with $i<j$ by usual transposition of the
monodromy matrix. On the other hand the formulas of the actions \eqref{act13}--\eqref{act23} are the
most simple among all the formulas given in section~\ref{sec:TijBV}.

Secondly, it worth mentioning that recently, some progress has been made in the calculation of scalar products involving on-shell BV
\cite{Whe12, BelPakRS12a, BelPakRS12c,BelPakRS12d}.  In particular, single determinant representations for form factors
of diagonal elements $T_{jj}(w)$ were obtained in \cite{BelPakRS12c,BelPakRS12d}, using a method based on the
twisted transfer matrix. It is naturally to expect that the same determinants for form factors can be derived directly
form the action of the operators $T_{jj}(w)$ on BV. The details of this derivation may provide the key to getting determinant representations for form factors of the operators $T_{ij}(w)$ with $i<j$. We hope to consider this problem in our further work.

\section*{Acknowledgements}
Work of S.P. was supported in part by RFBR grant 11-01-00962-a, grant
of Scientific Foundation of NRU HSE 12-09-0064 and grant of
FASI RF 14.740.11.0347. E.R. was supported by ANR Project
DIADEMS (Programme Blanc ANR SIMI1 2010-BLAN-0120-02).
N.A.S. was  supported by the Program of RAS Basic Problems of the Nonlinear Dynamics,
RFBR-11-01-00440, RFBR-11-01-12037-ofi-m, SS-4612.2012.1.

\appendix

\section{Properties of Izergin--Korepin determinant $\Izer_n(\bar x|\bar y)$\label{prop:Izer}}

The following properties of $\Izer_n$ are useful:
 \ben
 \label{Red-K}
\Izer_{n}(\bar x-c|\bar y) &=&
\Izer_{n}(\bar x|\bar y+c)= (-1)^n f^{-1}(\bar y,\bar x) \Izer_{n}(\bar y|\bar x)\,,
\\
 \label{K-K}
\Izer_{n+1}(\{\bar x, z-c\}|\{\bar y, z\}) &=& \Izer_{n+1}(\{\bar x, z\}|\{\bar y, z+c\})= - \Izer_{n}(\bar x|\bar y)\,,
\\
\Izer_{n}(\bar x|\bar y) &=& \Izer_{n}(-\bar y|-\bar x)
\mb{and} \Izer_{1}( x| y) = g(x,y),
\\
\Bigl.\Izer_{n}(\bar x|\bar y)\Bigr|_{x_n\to y_n}&= & g(x_n,y_n)
f(y_n,\bar y_{n})f(\bar x_{n},x_n) K_{n-1}(\bar x_{n}|\bar y_{n})+{\rm reg},\label{Resid-K}
%
%\\  f(\bar x;\bar x +c)\,\Izer_{n}(\bar x|\bar x) &=& 1
\een
where ${\rm reg}$ means the regular part, and we recall that $\bar x_{n}=\bar x\setminus x_n$ and
$\bar y_{n}=\bar y\setminus y_n$.
\begin{lemma}\label{main-ident}
Let $\bar\gamma$, $\bar\alpha$ and $\bar\beta$ be sets of complex variables with $\#\alpha=m_1$,
$\#\beta=m_2$, and $\#\gamma=m_1+m_2$. Then
\begin{equation}\label{Sym-Part-old1}
  \sum
 \Izer_{m_1}(\bar\gamma_{\so}|\bar \alpha)\Izer_{m_2}(\bar \beta|\bar\gamma_{\st})f(\bar\gamma_{\st},\bar\gamma_{\so})
 = (-1)^{m_1}f(\bar\gamma,\bar \alpha) \Izer_{m_1+m_2}(\{\bar \alpha-c,\bar \beta\}|\bar\gamma).
 \end{equation}
The sum is taken with respect to all partitions of the set $\bar\gamma\Rightarrow\{\bar\gamma_{\so},\bar\gamma_{\st}\}$ with $\#\bar\gamma_{\so}=m_1$ and $\#\bar\gamma_{\st}=m_2$.
Due to \eqref{Red-K} the equation \eqref{Sym-Part-old1} can be also written in the form
\begin{equation}\label{Sym-Part-old2}
  \sum
 \Izer_{m_1}(\bar\gamma_{\so}|\bar \alpha)\Izer_{m_2}(\bar \beta|\bar\gamma_{\st})f(\bar\gamma_{\st},\bar\gamma_{\so})
 = (-1)^{m_2}f(\bar \beta,\bar\gamma) \Izer_{m_1+m_2}(\bar\gamma|\{\bar \alpha,\bar \beta+c\}).
 \end{equation}
\end{lemma}
The proof of this lemma was given in \cite{BelPakRS12c}.

\section{Proofs of multiple exchange relations\label{prf:TT}}
%\subsection{Proof of proposition \ref{Lem1}}
The proof of relation \eqref{two} can be given by induction over $n_y=\# y$. For simplicity, we fixed $i=1$, $j=3$ and $k=2$, but obviously the calculation can be made whatever they are.
We start with $n_x=n_y=1$,
\beq{comm21}
T_{13}(y)T_{12}(x)=f(x,y)T_{12}(x)T_{13}(y)+g(y,x)T_{12}(y)T_{13}(x).
\eeq
Then the standard Bethe ansatz considerations allow us to generalize this formula to
the case $n_y=1$, $n_x\ge 1$,
\beq{comm21-nx}
T_{13}(y)T_{12}(\bar x)=f(\bar x,y)T_{12}(\bar x)T_{13}(y)+\sum g(y,\bar x_{\so})f(\bar x_{\st},\bar x_{\so})T_{12}(y)T_{12}(\bar x_{\st})T_{13}(\bar x_{\so}).
\eeq
Here sum is taken over the partitions $\bar x \Rightarrow \{\bar x_{\so},\bar x_{\st}\}$ with $\#\bar x_{\so}=1$.
 Comparing \eqref{comm21-nx} with \eqref{two} at $n_y=1$ we
see that the first term in \eqref{comm21-nx} corresponds to the partition $\bar w_{\so}=y$, $\bar w_{\st}=\bar x$, while
the second term corresponds to the partitions $\bar w_{\so}=\bar x_{\so}$, $\bar w_{\st}=\{\bar x_{\st},y\}$.

Then we proceed by induction over $n_y\equiv n$.
Due to the induction assumption and  \eqref{two-t} we have
 \beq{1step}
 T_{13}(\bar y)T_{12}(\bar x)=(-1)^{n-1}T_{13}(y_n)\sum  \Izer_{n-1}(\bar w_{\so}|\bar y_n+c)T_{12}(\bar w_{\st})T_{13}(\bar w_{\so})f(\bar w_{\st},\bar w_{\so}),
\eeq
where  $\{\bar y_n,\bar x\}=\bar w\Rightarrow\{\bar w_{\so},\bar w_{\st}\}$ with $\#\bar w_{\so}=n-1$, and $\bar y_n=\bar y\setminus y_n$. Now we should move $T_{13}(y_n)$ to the right, using
\eqref{two-t}  for $n_y=1$. Then we obtain
 \beq{2step}
 T_{13}(\bar y)T_{12}(\bar x)=(-1)^{n}\sum \Izer_1(\bar w_{\rm ii}|y_n+c) \Izer_{n}(\bar w_{\so}|\bar y_n+c)T_{12}(\bar w_{\rm i})T_{13}(\bar w_{\rm ii})T_{13}(\bar w_{\so})
 f(\bar w_{\st},\bar w_{\so}) f(\bar w_{\rm i},\bar w_{\rm ii}),
\eeq
where we have an additional sum over partitions $\{\bar w_{\st},y_n\}\Rightarrow \{\bar w_{\rm i}, \bar w_{\rm ii}\}$
with $\#\bar w_{\rm ii}=1$. Now we use
\beq{evid}
 \Izer_1(\bar w_{\rm ii}|y_n+c)f(\bar w_{\st},\bar w_{\so})= -\Izer_1(y_n|\bar w_{\rm ii})f(\bar w_{\rm i},\bar w_{\so})
 f(\bar w_{\rm ii},\bar w_{\so})f^{-1}(y_n,\bar w_{\so}) f^{-1}(y_n, \bar w_{\rm ii}).
 \eeq
Substituting this into \eqref{2step} and denoting $\{\bar w_{\rm ii},\bar w_{\so}\}=\bar w_0$ we obtain
 \beq{3step}
 T_{13}(\bar y)T_{12}(\bar x)=(-1)^{n+1}\sum \Bigl[\Izer_1(y_n|\bar w_{\rm ii})\Izer_{n-1}(\bar w_{\so}|\bar y_n+c)
 f(\bar w_{\rm ii},\bar w_{\so})\Bigr]T_{12}(\bar w_{\rm i})T_{13}(\bar w_0)
 \frac{f(\bar w_{\rm i},\bar w_0)}{f(y_n,\bar w_0)}.
\eeq
We  originally had only one restriction for the partitions: $y_n\notin \bar w_{\so}$. However now  we can consider the
partitions with $y_n\in \bar w_{\so}$ as well, since in this case the factor
$f^{-1}(y_n,\bar w_0)$ automatically vanishes.

It remains to apply \eqref{Sym-Part-old1} to the terms in the squared brackets. It gives
 \beq{4step}
 T_{13}(\bar y)T_{12}(\bar x)=(-1)^{n}\sum \Izer_{n}(\bar w_0|\bar y+c)T_{12}(\bar w_{\rm i})T_{13}(\bar w_0)
 f(\bar w_{\rm i},\bar w_0).
\eeq

Finally, relation \eqref{one} is obtained from \eqref{two} thanks to the morphism $\vph$.

\section{Proof of relation \eqref{eq:triangX} \label{app:X}}
We start with the defining relation for $X_j=
{\rm tr}_{\bar a_1,\bar b} \big(\ct_{\bar a_1,\bar b}(\bar\la_1;\bar\muu)\,E_j\big)\rvec$.
Using the RTT algebra we can show that
\ben \label{MTv}
\ct_{\wb{a}_1,\bar b}(\bar\la_1;\bar\muu)=\prod_{l=j+1}^bR_{B_lB_j}(v_l,v_j)
\ct_{\bar a_1\bar b_j}(\bar\la_1;\bar\muu_j)T_{B_j}(v_j) \prodl_{l=2}^aR_{B_jA_l}(v_j,u_l)  \prodl_{l=j+1}^bR_{B_jB_l}(v_j,v_l).
\een
The left product of $R$-matrices can be moved to the right using cyclicity of the traces, and then acts trivially on  $E_j$ from the right.
Then, $X_j$ can be rewritten as
\ben
X_j={\rm tr}_{\bar a_1,\bar b} \Big(\ct_{\bar a_1\bar b_j}(\bar\la_1;\bar\muu_j)T_{B_j}(v_j)
\prodl_{l=2}^aR_{B_jA_l}(v_j,u_l)  \prodl_{l=j+1}^bR_{B_jB_l}(v_j,v_l)\,E_j\Big).
\een
As before, we develop the action of $R$-matrices
\ben
 \prodl_{l=j+1}^bR_{B_jB_l}(v_j,v_l)\,E_j=\prod_{l=j+1}^bf^{-1}(v_j,v_l)E_j+\sum_{k=j+1}^bh^{-1}(v_j,v_k) \prod_{l=j+1}^{k-1}f^{-1}(v_j,v_l) E_k\,.
\een
The first term allows us to trace over the auxiliary space $B_j$ to get $\bbb^{a-1,b-1}(\bar\la_1;\bar\muu_j)$. In the second term we put back $T_{B_j}(v_j) $ in its original position using again (\ref{MTv}) and the unitary relation $R_{B_jB_i}(v_j,v_i)R_{B_iB_j}(v_i,v_j)=\mathbf{I}$. Then we obtain
\ben
X_j&=&\lambda_2(\bar v)\lambda_2(\bar u_1)\prod_{l=j+1}^bf^{-1}(v_j,v_l)\bbb^{a-1,b-1}(\bar\la_1;\bar\muu_j)\\
&&+\sum_{k=j+1}^bh^{-1}(v_j,v_k) \prod_{l=j+1}^{k-1}f^{-1}(v_j,v_l) {\rm tr}_{\bar a_1,\bar b} \big(\ct_{\bar a_1,\bar b}(\bar\la_1;\bar\muu)\prod_{l=j+1}^bR_{B_lB_j}(v_l,v_j)\,E_k\big).\nonumber
\een
Now we again  develop the action of the $R$-matrices on $E_k$ as
\ben
\prod_{l=j+1}^bR_{B_lB_j}(v_l,v_j)\,E_k
%&=&\prod_{l=j+1}^kR_{B_lB_j}(v_l,v_j)\,E_k^{a-1b}\\
&=&f^{-1}(v_k,v_j)E_k+h^{-1}(v_k,v_j)\prod_{l=j+1}^{k-1}f^{-1}(v_l,v_j)E_j\\
&&+h^{-1}(v_k,v_j)\sum_{i=j+1}^{k-1} h^{-1}(v_i,v_j)\prod_{l=i+1}^{k-1}f^{-1}(v_l,v_j)E_i\,.\nonumber
\een
It follows after some manipulation with the double sum that
\ben
G^{(j)}_{j}\,X_j+\sum_{k=j+1}^{b}I_{jk}\,G^{(j)}_{k}\,X_k&=&\lambda_2(\bar u_1)\lambda_2(\bar v)\prod_{l=j+1}^bf^{-1}(v_j,v_l)\bbb^{a-1,b-1}(\bar\la_1;\bar\muu_j),
\een
with
\ben
G^{(j)}_{k}&=&1-\sum_{i=k+1}^bh^{-1}(v_j,v_i)h^{-1}(v_i,v_j) \prod_{l=k+1}^{i-1}f^{-1}(v_j,v_l)f^{-1}(v_l,v_j),\qquad \\
I_{jk}&=&-h^{-1}(v_j,v_k) f^{-1}(v_k,v_j) \prod_{i=j+1}^{k-1}f^{-1}(v_j,v_i).
\label{eq:G1}
\een
One can then show by recursion on $b$ that
\ben
G^{(j)}_{k}=\prod_{l=k+1}^bf^{-1}(v_j,v_l)f^{-1}(v_l,v_j).
\label{eq:G2}
\een
The case $b=1$ is trivial. Then, considering $G^{(j)}_{k}$ as a function of $v_b$, one can easily see that the two expressions \eqref{eq:G1} and  \eqref{eq:G2} have same residues at all poles $v_b=v_\ell\pm c$. Moreover, equality of their limit $v_b\to\infty$ is just the induction hypothesis for $b-1$. Since the two expressions are rational functions of $v_b$, this proves that they are equal.

\medskip

Then
\ben
X_j+\sum_{k=j+1}^bg(v_k,v_j) \prod_{l=j+1}^{k-1}f(v_l,v_j)\,X_k=\lambda_2(\bar u_1)\lambda_2(\bar v)\prod_{l=j+1}^{b}f(v_l,v_j)\bbb^{a-1,b-1}(\bar\la_1;\bar\muu_j).
\een

%%%%%%%%%%%%%%%%%%%%%%%%%%%%%%%%%%%

\end{document}